\documentclass[10pt ]{revtex4}
\usepackage{hyperref}
\usepackage{amsmath,amsfonts}
\usepackage[dvips]{graphics}
\usepackage[dvips]{graphicx}
\usepackage{subfigure}
\usepackage{epsfig}
\usepackage{amsmath}
\usepackage{amsfonts}

\begin{document}

\title{Thermal fluctuations of charged black holes in gravity's rainbow}
\author{Sudhaker Upadhyay${}^{a,b}$}
\email{sudhakerupadhyay@gmail.com; sudhaker@associates.iucaa.in}
\author{ Seyed Hossein Hendi${}^{c,d}$}
\email{hendi@shirazu.ac.ir}
\author{Shahram Panahiyan${}^{c,e,f}$}
\email{shahram.panahiyan@uni-jena.de}
\author{Behzad Eslam Panah${}^{c,g}$}
\email{behzad.eslampanah@gmail.com}
\affiliation{$^{a}$ Department of Physics, K.L.S. College Nawada, Nawada-805110, Bihar,
India}
\affiliation{${}^{b}$ Inter-University Centre for Astronomy
and Astrophysics (IUCAA) Pune, Maharashtra-411007}
\affiliation{$^{c}$ Physics Department and Biruni Observatory, College of Sciences,
Shiraz University, Shiraz 71454, Iran}
\affiliation{$^{d}$ Research Institute for Astronomy and Astrophysics of Maragha (RIAAM),
P.O. Box 55134-441, Maragha, Iran}
\affiliation{$^{e}$ Helmholtz-Institut Jena, Fr\"{o}belstieg 3, Jena 07743, Germany}
\affiliation{$^{f}$ Physics Department, Shahid Beheshti University, Tehran 19839, Iran}
\affiliation{$^{g}$ ICRANet, Piazza della Repubblica 10, I-65122 Pescara, Italy}

\begin{abstract}
Quantum fluctuation effects have an irrefutable role in high
energy physics. Such fluctuation can be often regarded as a
correction of infrared (IR) limit. In this paper, the effects of
the first-order correction of entropy, caused by thermal
fluctuation, on the thermodynamics of charged black holes in
gravity's rainbow will be discussed. It will be shown that such
correction has profound contributions to high energy limit of
thermodynamical quantities, stability conditions of the black
holes and interestingly has no effect on thermodynamical phase
transitions. The coupling between gravity's rainbow and the
first-order correction will be addressed. In addition, the
measurement of entropy as a function of fluctuation of temperature
will be done and it will be shown that de Sitter (dS) case
enforces an upper limit on the values of temperature and produces
cyclic like diagrams. While for the anti-de Sitter (AdS) case, a
lower limit on the entropy is provided and although for special
cases a cyclic like behavior could be observed, no upper or lower
limit exists for the temperature. In addition, a comparison
between non-correction and correction included cases on the
thermodynamical properties of solutions will also be discussed and
the effects of the first-order correction will be highlighted. It
will be shown that the first-order correction provides the
solutions with larger classes of thermal stability conditions
which may result into existence of a larger number of
thermodynamical structures for the black holes.
\end{abstract}

\maketitle

\section{Introduction}

Finding a consistent quantum theory of gravity is one of the interesting
subjects in physics. Accordingly, a lot of attempts to join gravity and
quantum theories together have been done. Nonetheless, there is no complete
description of the quantum gravity yet. It is notable that, in order to
construct a quantum theory of gravity, we have to investigate the validation
of theory in the ultraviolet (UV) regime. On the other hand, it is arguable
that Einstein's theory of general relativity is an effective theory which is
valid in IR limit, while in UV regime it fails to produce accurate results.
Therefore, this shortcoming requires modification in order to incorporate
the UV regime. Horava-Lifshitz gravity is one of these theories which
includes the UV regime by considering a modification of the usual
energy-momentum relation \cite{HoravaI,HoravaII}. This theory reduces to
general relativity (GR) in the IR limit. In other words, we can consider
this theory as a UV completion of GR. Motivated by this work on
Horava-Lifshitz gravity, different Lifshitz scaling for space and time have
been considered for various theories, namely, type IIA string theory \cite%
{Gregory}, type IIB string theory \cite{Burda}, dilatonic black branes \cite%
{Goldstein,Bertoldi}, dilatonic black holes \cite{Kord,Tarrio}, and AdS/CFT
correspondence \cite{Gubser,Ong,Alishahiha,Kachru}.

It is notable that similar to the Horava-Lifshitz gravity, the
gravity's rainbow (doubly GR) \cite{MagueijoS2002,MagueijoS2004},
has also been viewed as a UV completion of GR. In doubly special
relativity, there are two fundamental constants; (i) the velocity
of light, (ii) the Planck energy. In this theory, a particle
cannot attain velocity and energy larger than the velocity of
light and the Planck energy, respectively. Gravity's rainbow is a
generalization of doubly special relativity to curved spacetime.
In this theory of gravity, the energy of test particle affects the
geometry of spacetime. In other words, gravity has different
effects on the particles with various energies. So, the geometry
of spacetime is represented by a family of energy-dependent
metrics forming a rainbow of metrics. Similar to Horava-Lifshitz
gravity, the gravity's rainbow can be constructed by considering a
modification of the standard energy-momentum relation. The
modification of usual energy-momentum relation in this gravity
theory is given as
\begin{equation}
E^{2}f^{2}\left( \varepsilon \right) -p^{2}g^{2}\left( \varepsilon \right)
=m^{2},
\end{equation}%
with $\varepsilon =E/E_{p}$, where $E$ and $E_{p}$ are the energy of a test
particle and the Planck energy, respectively. The functions $f\left(
\varepsilon \right) $ and $g\left( \varepsilon \right) $ are called rainbow
functions in which they are phenomenologically motivated and could be
extracted by experimental data \cite{AliK2015}. The rainbow functions are
required to satisfy $\underset{\varepsilon \rightarrow 0}{\lim }f\left(
\varepsilon \right) =1$ and $\underset{\varepsilon \rightarrow 0}{\lim }%
g\left( \varepsilon \right) =1$, where these conditions ensure
that modified energy-momentum relation reduces to the usual one in
the IR limit. It is worthwhile to mention that the energy of a
test particle ($E$) cannot be greater than $E_{p}$. It means that
if a test particle is used to probe the geometry of spacetime,
then $E$ is always smaller than $E_{p}$ (see Ref. \cite{Peng}, for
more details). It is notable that such justification is based on
the Standard model of particle physics. Also, considering a
suitable choice of the rainbow functions, the Horava-Lifshitz
gravity can be related to the gravity's rainbow \cite{Garattini}.
It should be noted that since it was shown that specific models of gravity's rainbow could be related to Horava-Lifshitz gravity  \cite{Garattini}, one can argue that, by suitable transformations, the thermal effects obtained and investigated here, could be mapped  to Horava-Lifshitz results with specific scaling parameter. This enables one to perform study in less complex framework and map it later to a more sophisticated one (similar to the AdS/CFT correspondence approach).

One of the most important results of the UV completion of
geometries is modification at the last stage of evaporation of
black holes. In the IR limit where gravity's rainbow yields
general relativity, the thermodynamics of black holes in gravity's
rainbow reduces to the usual one which is observed for large black
holes. On the contrary, in case of the black holes evaporation and
reduction in their size, the black hole thermodynamics in
gravity's rainbow shows significant deviation from the usual black
hole thermodynamics. Especially, such deviation becomes more
evident at the end stage of evaporation of black hole in gravity's
rainbow \cite{Ali}. At this stage, the temperature of black holes
acquires a maximum value, and then it decreases beyond this
maximum value. At a critical point, while size of the black hole
is non-zero, its temperature vanishes and, therefore, there is no
Hawking radiation. This indicates the existence of a black hole
remnant in gravity's rainbow. Formation and existence of a black
hole remnant play a crucial role in phenomenological consequences
for the detection of mini black holes at the LHC \cite{Ali}. In
order to incorporate the upper bound on the energy of a particle,
the usual uncertainty principle has to be modified to a
generalized one. The particle probes the geometry of black hole
and it also fixes the energy scale of the gravity's rainbow. This
provides the possibility of using the bound on the energy of a
particle emitted in the Hawking radiation as an energy scale in
the rainbow functions.

Black hole solutions and their thermodynamical properties have
been obtained in gravity's rainbow, and it was shown that the
rainbow functions modify the
thermodynamical properties of obtained black holes in this theory \cite%
{Ali,Galan,Ling,PengWu,LiLing,Gim,HendiPEMEPJC,Kim,Feng,FengII}.
Also effects of gravity's rainbow on the thermodynamical
properties and phase
transition of black holes have been investigated in Gauss-Bonnet \cite%
{GBRainbowII}, Kaluza-Klein \cite{Alsaleh}, massive
\cite{MassRainII,suhe}, F(R) \cite{F(R)I,F(R)}, and dilaton
\cite{dilatonI,dilatonII} gravities. Electric charge and magnetic
monopoles in gravity's rainbow have been studied in Ref.
\cite{GarattiniM}. In the context of gravity's rainbow, the
possibility of quantum fluctuations to induce a topological change
has been studied \cite{GarattiniL}. Using the gravity's rainbow,
the hydrostatic equilibrium for compact objects and the structure
of neutron stars have also been investigated in Refs.
\cite{TOVI,TOVII}. The effects of rainbow function on wormholes
have been studied in Ref. \cite{Worm}. Cosmic string
and black rings in gravity's rainbow have been obtained in Refs. \cite%
{Momeni} and \cite{AliFKh}, respectively. Metric structures of
this theory such as; the original rainbow metrics proposal, the
momentum-space-inspired metric and the standard Finsler geometry
approach have been perused in Ref. \cite{Lobo}. The effects of
rainbow functions on the geodesics around the black holes have
been studied in Ref.\cite{Leiva}. Another interesting result of
this gravity is related to the FRW cosmology in which it was shown
that employing this formalism will provide the possibility of
removing the big
bang singularity \cite%
{NonsingularI,NonsingularII,NonsingularIII,NonsingularIV}, and the big
bounce of a cyclic universe \cite{LingWu}. For more investigations regarding
the gravity's rainbow, see Refs. \cite%
{MotivI,MotivII,MotivIII,MotivIV,MotivV,MotivVI,MotivVII,MotivVIII,MotivIX,MotivX,MotivXI}%
.

Moreover, it has been observed that thermodynamics of black hole
undergoes corrections due to the thermal fluctuations. The various
approaches, namely, field theory methods \cite{80,90}, quantum
geometry techniques \cite{110,120}, general statistical mechanical
arguments \cite{l1} and Cardy formula \cite{carl} had confirmed
that leading-order corrections to the entropy originated due to
the thermal fluctuations are logarithmic in nature. The
consequences of these corrections on the various black hole
thermodynamics have been studied with great interests \cite{sud10,sud11,sud12,sud13}. For
instance, recently, the role of thermal fluctuations  has been
studied in the contexts of  Schwarzschild-Beltrami-de Sitter black
hole \cite{sud2} and the massive black hole in AdS space
\cite{sud3}. Similar analysis has also been made on the modified
Hayward black hole, where it is found that thermal fluctuation
plays an important role in order to reduce pressure and internal
energy of the Hayward black hole \cite{behn}. The importance of
logarithmic correction to the area law can also be seen in the
study of quark-gluon plasma properties through holographic
principles \cite{beh1,beh2}.  Our motivation, here, is to study
the effects of thermal fluctuations on the thermodynamics of
charged black hole in gravity's rainbow.

The paper is presented as follows. In Sec. II, we discuss the
charged black hole in gravity's rainbow. We study the effects of
thermal fluctuation on various thermodynamical quantities in Sec.
III. The effects of the first-order correction on thermal
stability are discussed in Sec. IV. Section V devoted to
investigate the correction case versus the non-correction one on
thermodynamical properties of the solutions. We, finally, conclude
our results in the last section.

\section{ Charged Black holes in gravity's rainbow}

The Lagrangian density of Einstein gravity with cosmological constant
coupled to an electromagnetic field is given by
\begin{equation}
L=R-2\Lambda -L_{Max},
\end{equation}%
where $L_{Max}=F_{ab}F^{ab}$ is the Maxwell's Lagrangian.

The equations of motion corresponding to the metric tensor $g_{ab}$ and the
Faraday tensor $F_{ab}$ are respectively
\begin{eqnarray}
R_{ab}+g_{ab}\left( \Lambda \left( \varepsilon \right) -\frac{1}{2}R\right)
+2\left( F_{ac}F_{b}^{c}-\frac{1}{4}g_{ab}F_{cd}F^{cd}\right) &=&0,
\label{eq} \\
\partial _{a}(\sqrt{-g}F^{ab}) &=&0,  \label{eq1}
\end{eqnarray}%
where $\Lambda \left( \varepsilon \right) $ is the energy-dependent
cosmological constant. The energy-dependent metric for gravity's rainbow is
defined by
\begin{equation}
ds^{2}=-\frac{\psi (r,\varepsilon )}{f^{2}(\varepsilon )}+\frac{dr^{2}}{%
g^{2}(\varepsilon )\psi (r,\varepsilon )}+\frac{r^{2}}{g^{2}(\varepsilon )}%
h_{ij}dx^{i}dx^{j},\ i,j=1,2.
\end{equation}%
where $h_{ij}dx^{i}dx^{j}$ is the spacial line element with constant
curvature $2k$. The incorporation of gauge potential ansatz $%
A_{a}=h(r,\varepsilon )\delta _{a}^{0}$ in the Maxwell equation (\ref{eq1})
yields
\begin{equation}
h(r,\varepsilon )=-\frac{q(\varepsilon )}{r},
\end{equation}%
where $q\left( \varepsilon \right) $ is as integration constant related to
the electrical charge.  Here, the electric charge is an invariant
quantity. We know that $q\left( \varepsilon \right) $ is an integration constant which is proportional
to electric charge. It is notable that the integration constant can be energy
dependent, too. In order to clarify this point, we can regard $q(\epsilon)=g(\epsilon)^a f(\epsilon)^b q$ with arbitrary $a$ and $b$ and $q$ is the invariant electric charge.  Also, it
leads to the following nonzero component of electromagnetic field tensor $%
F_{rt}=\frac{q(\varepsilon )}{r^{2}}$. The equation of motion (\ref{eq})
leads to
\begin{equation}
\psi ^{\prime }(r,\varepsilon )g^{2}(\varepsilon )r+(\psi (r,\varepsilon
)-k)g^{2}(\varepsilon )+\Lambda (\varepsilon )r^{2}+\frac{q^{2}(\varepsilon
)g^{2}(\varepsilon )f^{2}(\varepsilon )}{r^{2}}=0,
\end{equation}%
which has the following solution:
\begin{equation}
\psi (r,\varepsilon )=k-\frac{m_{0}(\varepsilon )}{r}-\frac{\Lambda
(\varepsilon )r^{2}}{3g^{2}(\varepsilon )}+\frac{q^{2}(\varepsilon
)f^{2}(\varepsilon )}{r^{2}},  \label{metric function}
\end{equation}%
where $m_{0}(\varepsilon )$ is an integration constant related to the total
mass of the black hole.  We note that, in the GR limit (i.e., $f(\epsilon)\rightarrow 1$ and $g(\epsilon)\rightarrow 1 $), the solution (\ref{metric function}) reduces to the
 Reissner-Nordstr\"om black hole
in 4-dimensions. Here, one may take $k=1,0,-1$, corresponding to a
spherical, Ricci flat, hyperbolic horizon for the black hole, respectively.

\section{Thermodynamical quantities}

It is possible to obtain the Hawking temperature by using the surface
gravity as follows,
\begin{eqnarray}
T_{H} &=&\frac{g(\varepsilon )}{4\pi f(\varepsilon )}\psi ^{\prime
}(r,\varepsilon )|_{r=r_{+}},  \notag \\
&=&\frac{1}{4\pi }\left( \frac{kg(\varepsilon )}{f(\varepsilon )r_{+}}-\frac{%
\Lambda (\varepsilon )r_{+}}{g(\varepsilon )f(\varepsilon )}-\frac{%
q^{2}(\varepsilon )g(\varepsilon )f(\varepsilon )}{r_{+}^{3}}\right) ,
\label{tem}
\end{eqnarray}%
in which $r_{+}$ is related to the event horizon of black hole. it can be seen here that
in the GR limit (i.e., $f(\epsilon)\rightarrow 1$ and $g(\epsilon)\rightarrow 1 $), 
we recover the  Hawking temperature of charged black holes. 
  As we know
that the thermal fluctuations correct thermodynamical quantities. To the
leading order, the entropy gets logarithmic correction \cite{l1,SPR} and
given by,
\begin{equation}
S=S_{0}-\frac{\alpha }{2}\log (S_{0}T_{H}^{2}),  \label{entropy}
\end{equation}%
where $\alpha $ is a dimensional full parameter and the zeroth-order entropy
in four-dimensions is given by \cite{HendiPEMEPJC}
\begin{equation}
S_{0}=\frac{\pi r_{+}^{2}}{4g^{2}(\varepsilon )}.  \label{ent}
\end{equation}

Exploiting relations (\ref{tem}), (\ref{entropy}) and (\ref{ent}), we get
explicit expression of first-order corrected entropy as
\begin{equation}
S=\frac{\pi r_{+}^{2}}{4g^{2}(\varepsilon )}-{\alpha }\log \left[ \frac{1}{8%
\sqrt{\pi }}\left( \frac{k}{f(\varepsilon )}-\frac{\Lambda (\varepsilon
)r_{+}^{2}}{g^{2}(\varepsilon )f(\varepsilon )}-\frac{q^{2}(\varepsilon
)f(\varepsilon )}{r_{+}^{2}}\right) \right] .  \label{en}
\end{equation}

Utilizing the definitions of entropy and temperature, we are able to compute
the Helmholtz function with following relation: $F=-\int SdT_{H}$. Here, the
Helmholtz function is calculated by
\begin{eqnarray}
F &=&\frac{1}{16}\left( \frac{kr_{+}}{f(\varepsilon )g(\varepsilon )}+\frac{1%
}{3}\frac{\Lambda (\varepsilon )r_{+}^{3}}{g^{3}(\varepsilon )f(\varepsilon )%
}+\frac{3q^{2}(\varepsilon )f(\varepsilon )}{g(\varepsilon )r_{+}}\right) +%
\frac{\alpha }{6\pi }\frac{q^{2}(\varepsilon )f(\varepsilon )g(\varepsilon )%
}{r_{+}^{3}}  \notag \\
&+&\frac{\alpha }{2\pi }\frac{\Lambda (\varepsilon )r_{+}}{f(\varepsilon
)g(\varepsilon )}-\frac{\alpha }{4\pi }\left( \frac{q^{2}(\varepsilon
)f(\varepsilon )g(\varepsilon )}{r_{+}^{3}}-\frac{kg(\varepsilon )}{%
f(\varepsilon )r_{+}}+\frac{\Lambda (\varepsilon )r_{+}}{f(\varepsilon
)g(\varepsilon )}\right) \log \left[ \frac{1}{8\sqrt{\pi }}\left( \frac{k}{%
f(\varepsilon )}\right. \right.  \notag \\
&-&\left. \left. \frac{\Lambda (\varepsilon )r_{+}^{2}}{g^{2}(\varepsilon
)f(\varepsilon )}-\frac{q^{2}(\varepsilon )f(\varepsilon )}{r_{+}^{2}}%
\right) \right] .  \label{Helmholtz}
\end{eqnarray}

Using the Hamiltonian approach, one can find the uncorrected mass $M$ of the
black hole for gravity's rainbow as \cite{MassRainII}
\begin{equation}
M=\frac{m_{0}}{8f(\varepsilon )g(\varepsilon )},  \label{ma}
\end{equation}%
where $m_{0}$ is evaluated by setting metric function to zero on horizon ($%
\psi (r,\varepsilon )|_{r=r+}=0$) as
\begin{equation}
m_{0}=kr_{+}-\frac{\Lambda (\varepsilon )r_{+}^{3}}{3g^{2}(\varepsilon )}+%
\frac{q^{2}(\varepsilon )f^{2}(\varepsilon )}{r_{+}}.
\end{equation}%
Here we note that the mass calculated here is not a corrected
mass. In other words, using this mass with first law of black hole
thermodynamics does not yield desirable result. To solve this
problem, we use thermodynamical approach in calculating mass. In
order to calculate the corrected mass $M$, we employ the concepts
of thermodynamics and use the following thermodynamical relation:
\begin{eqnarray}
M &=&F+T_{H}S,  \notag \\
&=&\frac{1}{8}\left( \frac{kr_{+}}{f(\varepsilon )g(\varepsilon )}-\frac{1}{3%
}\frac{\Lambda (\varepsilon )r_{+}^{3}}{g^{3}(\varepsilon )f(\varepsilon )}+%
\frac{q^{2}(\varepsilon )f(\varepsilon )}{g(\varepsilon )r_{+}}\right) +%
\frac{\alpha }{6\pi }\frac{q^{2}(\varepsilon )f(\varepsilon )g(\varepsilon )%
}{r_{+}^{3}}  \notag \\
&+&\frac{\alpha }{2\pi }\frac{\Lambda (\varepsilon )r_{+}}{f(\varepsilon
)g(\varepsilon )}.  \label{mass}
\end{eqnarray}

Obtained mass here could be used alongside of first law of black hole
thermodynamics to ensure the validity of calculated thermodynamical
quantities. The electric charge is calculated from the flux of the electric
field at infinity as follows,
\begin{equation}
Q=\frac{q(\varepsilon )f(\varepsilon )}{4g(\varepsilon )}.  \label{ch}
\end{equation}

The electric (chemical) potential $U$, can be defined as the gauge potential
at the event horizon with respect to the reference
\begin{equation}
U=A_{\mu }\chi ^{\mu }\left\vert _{r\rightarrow \infty }\right. -A_{\mu
}\chi ^{\mu }\left\vert _{r\rightarrow r_{+}}\right. =\frac{q(\varepsilon )}{%
r_{+}}.  \label{el}
\end{equation}

The first-order corrected Gibbs free energy for this black holes is given by
\begin{eqnarray}
G &=&F-UQ,  \notag \\
&=&\frac{1}{16}\left( \frac{k}{f(\varepsilon )g(\varepsilon )}r_{+}+\frac{1}{%
3}\frac{\Lambda (\varepsilon )r_{+}^{3}}{g^{3}(\varepsilon )f(\varepsilon )}-%
\frac{q^{2}(\varepsilon )f(\varepsilon )}{g(\varepsilon )r_{+}}\right) +%
\frac{\alpha }{6\pi }\frac{q^{2}(\varepsilon )f(\varepsilon )g(\varepsilon )%
}{r_{+}^{3}}  \notag \\
&+&\frac{\alpha }{2\pi }\frac{\Lambda (\varepsilon )r_{+}}{f(\varepsilon
)g(\varepsilon )}-\frac{\alpha }{4\pi }\left( \frac{q^{2}(\varepsilon
)f(\varepsilon )g(\varepsilon )}{r_{+}^{3}}-\frac{kg(\varepsilon )}{%
f(\varepsilon )r_{+}}+\frac{\Lambda (\varepsilon )r_{+}}{f(\varepsilon
)g(\varepsilon )}\right) \log \left[ \frac{1}{8\sqrt{\pi }}\left( \frac{k}{%
f(\varepsilon )}\right. \right.  \notag \\
&-&\left. \left. \frac{\Lambda (\varepsilon )r_{+}^{2}}{g^{2}(\varepsilon
)f(\varepsilon )}-\frac{q^{2}(\varepsilon )f(\varepsilon )}{r_{+}^{2}}%
\right) \right] .  \label{Gibbs}
\end{eqnarray}

Now, it is easy to establish the first law of black hole by using
thermodynamical quantities such as temperature (\ref{tem}), entropy (\ref{en}%
), charge (\ref{ch}), electric potential (\ref{el}) and corrected mass (\ref%
{mass}),
\begin{equation*}
dM=T_{H}dS+UdQ.
\end{equation*}

Our next quantity of interest is heat capacity. The importance of this
quantity lies in the fact that condition for thermal stability and phase
transition points could be extracted using this quantity. The heat capacity
of these black holes could be calculated as
\begin{eqnarray}
C &=&\frac{T_{H}}{\left( \frac{\partial ^{2}M}{\partial S^{2}}\right) _{Q}}=%
\frac{T_{H}}{{\left( \frac{\partial T_{H}}{\partial S}\right) }_{Q}}=\frac{%
\partial M}{\partial T_{H}},  \notag \\
&=&-\frac{\pi r_{+}^{2}}{2g(\varepsilon )}\left( \frac{kg^{2}(\varepsilon
)r_{+}^{2}-\Lambda (\varepsilon )r_{+}^{4}-q^{2}(\varepsilon
)g^{2}(\varepsilon )f^{2}(\varepsilon )}{kg^{2}(\varepsilon
)r_{+}^{2}+\Lambda (\varepsilon )r_{+}^{4}-3q^{2}(\varepsilon
)g^{2}(\varepsilon )f^{2}(\varepsilon )}\right)  \notag \\
&+&2\alpha \left( \frac{q^{2}(\varepsilon )f^{2}(\varepsilon
)g^{2}(\varepsilon )-\Lambda (\varepsilon )r_{+}^{4}}{kg^{2}(\varepsilon
)r_{+}^{2}+\Lambda (\varepsilon )r_{+}^{4}-3q^{2}(\varepsilon
)g^{2}(\varepsilon )f^{2}(\varepsilon )}\right) .  \label{heat}
\end{eqnarray}

In next section, we will investigate the thermodynamical behavior of the
solutions and highlight the effects of entropy correction in details.

\section{ Thermal stability}

Our main motivation here is to understand how the first-order correction
would modify thermodynamical behavior of the solutions. First, we start with
mass. It was seen that uncorrected version of the mass (\ref{ma}) does not
yield suitable results considering the concept of the first law of black
hole thermodynamics. Therefore, we calculated the corrected version using
thermodynamical concept (\ref{mass}).

Evidently, the corrected mass is modified due to presence of the first-order
correction. The mass here is an increasing function of the correction
parameter, $\alpha $. In addition, due to presence of the correction, two
new terms are added to mass. These two terms ($\frac{\alpha }{6\pi }\frac{%
q^{2}(\varepsilon )f(\varepsilon )g(\varepsilon )}{r_{+}^{3}}$\ and $\frac{%
\alpha }{2\pi }\frac{\Lambda (\varepsilon )r_{+}}{f(\varepsilon
)g(\varepsilon )}$ of (\ref{mass})) include electric charge coming from the
matter field contribution and cosmological constant which could be
understood as a generalization of the gravitational sector. Interestingly,
there is no coupling between the topological factor, $k$, and correction
parameter, $\alpha $. This signals the fact that first-order correction
affects the gravitational and matter field sectors while it does not have
any effect on topological structure. In addition, the thermodynamical
behavior of the mass, hence internal energy, in high energy regime (very
small $r_{+}$) is governed by the term in which electric charge and
correction parameter are coupled (term $\frac{\alpha }{6\pi }\frac{%
q^{2}(\varepsilon )f(\varepsilon )g(\varepsilon )}{r_{+}^{3}}$ in Eq. (\ref%
{mass})). This indicates that for high energy limit, the contribution of the
first-order correction is dominant and highlighted. On the other hand, the
asymptotical behavior of the mass is governed by $\Lambda $ term without
correction which shows the absence of effects of the first-order correction
for this limit (term $\frac{-1}{24}\frac{\Lambda (\varepsilon )r_{+}^{3}}{%
g^{3}(\varepsilon )f(\varepsilon )}$ in Eq. (\ref{mass})). Interestingly,
the topological term and correction term coupled with cosmological constant
have same order of horizon radius. By suitable choices of different
parameter ($k=-\frac{4\alpha }{\pi }\Lambda (E)$), it is possible to omit
the effects of topological term. This indicates that by this choice, one can
eliminate the effects of topological structure on thermodynamical behavior
of the mass. In other words, total behavior of the mass would be independent
of the geometrical structure of horizon (hence, topological factor). On the
other hand, due to this property, one can state that medium class of the
black holes are also affected by the existence of first-order correction. To
summarize, one can state that small and medium black holes are affected by
the presence of first-order correction while large black holes are not
affected on significant level.

In classical thermodynamics of the black holes, negativity of the internal
energy, hence mass signals the absence of physical black holes. It is a
matter of calculation to obtain the root for mass of this black hole in the
following form:
\begin{equation}
r_{+}\left( M=0\right) =\sqrt{\frac{g\left( \varepsilon \right) }{2\pi
\Lambda \left( \varepsilon \right) }\left[ \mathcal{A}^{1/3}+\frac{4\mathcal{%
B}}{\mathcal{C}^{1/3}}+2g\left( \varepsilon \right) \left( \pi k+4\alpha
\Lambda \left( \varepsilon \right) \right) \right] },  \label{RootM}
\end{equation}%
in which $\mathcal{A}$, $\mathcal{B}$ and $\mathcal{C}$ have following
forms:
\begin{eqnarray*}
\mathcal{A} &=&96\alpha g\left( \varepsilon \right) \Lambda \left(
\varepsilon \right) \left\{ g^{2}(\varepsilon )\left[ \frac{16}{3}\alpha
^{2}\Lambda ^{2}\left( \varepsilon \right) +4\pi k\alpha \Lambda \left(
\varepsilon \right) +\pi ^{2}k^{2}\right] +\frac{2\pi }{3}q^{2}(\varepsilon
)f^{2}(\varepsilon )\Lambda \left( \varepsilon \right) \right\} \\
&&+4\pi ^{3}kg\left( \varepsilon \right) \left[ 3q^{2}(\varepsilon
)f^{2}(\varepsilon )\Lambda \left( \varepsilon \right)
+2k^{2}g^{2}(\varepsilon )\right] +\mathcal{G}, \\
&& \\
\mathcal{B} &=&g^{2}(\varepsilon )\left[ \pi k+4\alpha \Lambda \left(
\varepsilon \right) \right] ^{2}+\pi ^{2}q^{2}(\varepsilon
)f^{2}(\varepsilon )\Lambda \left( \varepsilon \right) , \\
&& \\
\mathcal{C} &=&8g\left( \varepsilon \right) \left\{ g^{2}(\varepsilon )\left[
\pi k+4\alpha \Lambda \left( \varepsilon \right) \right] ^{3}+\frac{3\pi ^{2}%
}{2}q^{2}(\varepsilon )f^{2}(\varepsilon )\Lambda \left( \varepsilon \right) %
\left[ \pi k+\frac{16}{3}\alpha \Lambda \left( \varepsilon \right) \right]
\right\} +\mathcal{G}, \\
\mbox{with} \\
\mathcal{G} &=&4\pi q(\varepsilon )f(\varepsilon )\Lambda \left( \varepsilon
\right) \left[ 16\alpha g^{4}(\varepsilon )\left[ \pi k+4\alpha \Lambda
\left( \varepsilon \right) \right] ^{3}\right. \\
&-&\left. 3\pi ^{2}q^{2}(\varepsilon )g^{2}(\varepsilon )f^{2}(\varepsilon )
\left[ \pi ^{2}k^{2}-\frac{64}{3}\alpha ^{2}\Lambda ^{2}\left( \varepsilon
\right) \right] -4\pi ^{4}q^{4}(\varepsilon )f^{4}(\varepsilon )\Lambda
\left( \varepsilon \right) \right] ^{1/2}.
\end{eqnarray*}

\begin{figure}[tbp]
$%
\begin{array}{cc}
\epsfxsize=7cm \epsffile{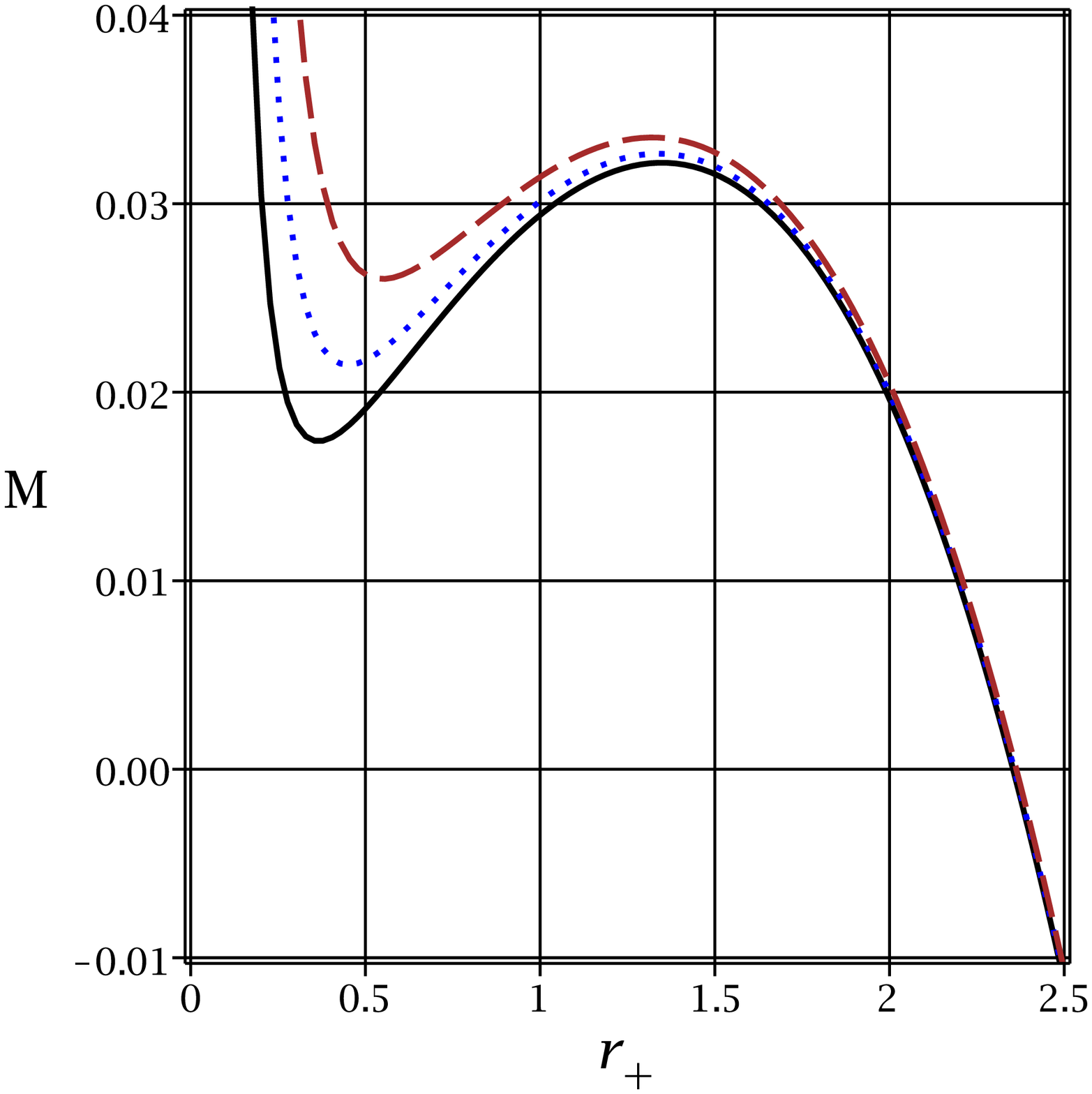} & \epsfxsize=7cm \epsffile{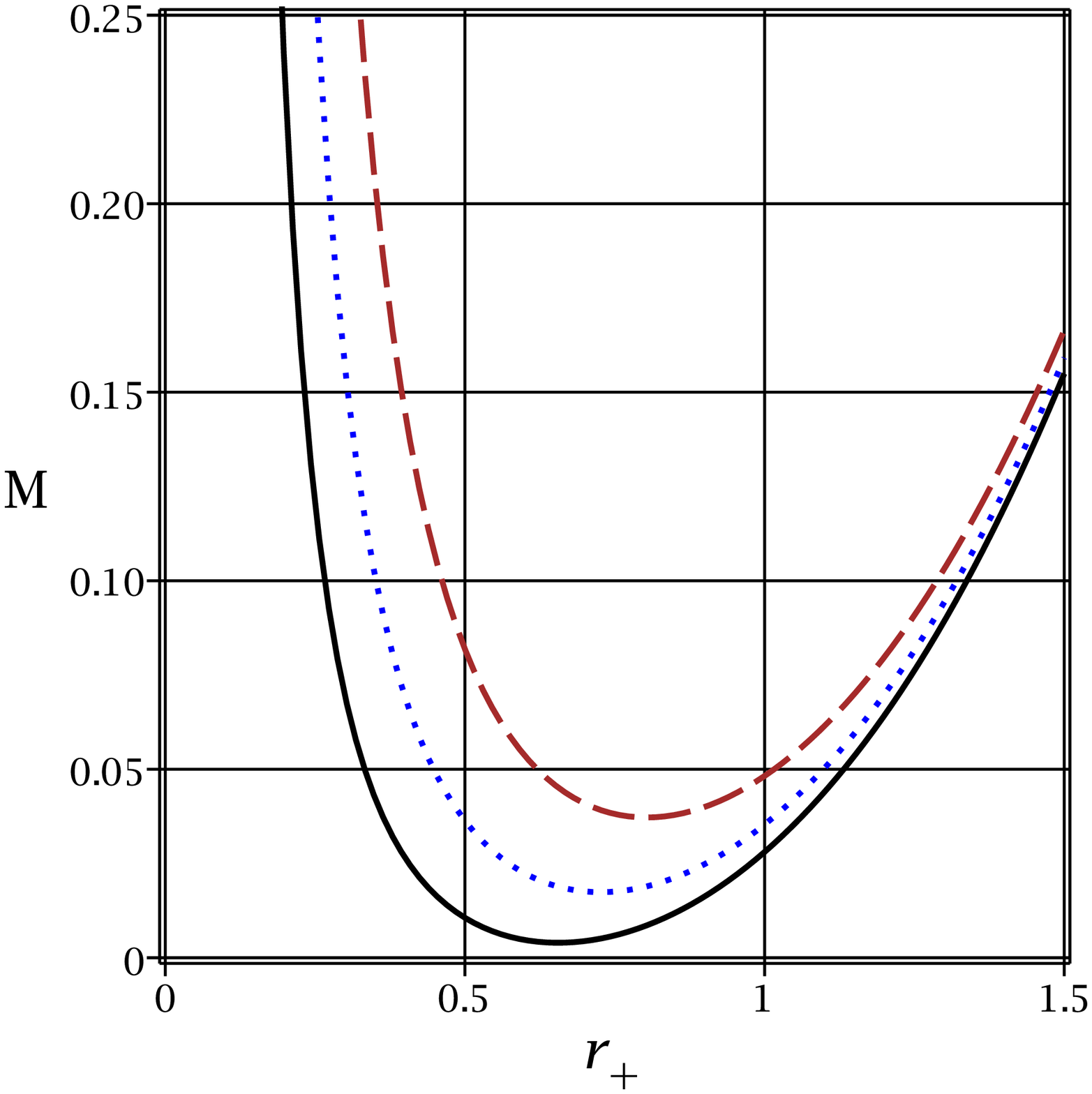}%
\end{array}
$%
\caption{$M$ versus $r_{+}$ for $f(\protect\varepsilon )=g(\protect%
\varepsilon )=0.9$, $k=\protect\alpha =1$, $q(\protect\varepsilon )=0.2$
(continuous line), $q(\protect\varepsilon )=0.2887$ (dotted line) and $q(%
\protect\varepsilon )=0.4$ (dashed line). \newline
Left panel: $\Lambda (\protect\varepsilon )=1$; Right panel: $\Lambda (%
\protect\varepsilon )=-1$.}
\label{FigM}
\end{figure}

Studying the root shows that the physical region for the existence
of the black hole is a function of the correction. In other words,
depending on choices of this parameter, the physical region for
the existence of black holes would be modified. In order to
understand the details regarding the behavior of the mass, we have
plotted two sets of diagrams for cases of dS (left panel of Fig.
\ref{FigM}) and AdS black holes (right panel of Fig. \ref{FigM}).
Evidently, for AdS case, mass enjoys a minimum in its structure
and it does not acquire root. On the other hand, dS case enjoys
one minimum, one maximum and one root in its structure. The root
is located after the extrema. After root, the mass is negative
which indicates that existence of black holes in dS picture is
limited to a specific region while for the AdS case, such limit
does not exist. The place of the root is an increasing function of
the correction parameter. This means that comparing to uncorrected
version, in the presence of first-order correction, the region of
positivity of mass, hence internal energy, is bigger which
indicates that physical black holes with larger horizon are
obtainable in the presence of this correction.

Next, we turn our attention to temperature. The temperature is
independent of the first-order correction. This means that
although that entropy and mass of the black holes are affected by
the presence of first-order correction, the temperature is free of
it. This shows that presence of the first-order correction
modified the thermodynamical structure of these black holes on a
specific level and introduced the following property: the entropy
is a decreasing function of the correction parameter. This means
that by increasing this parameter, the disorder which is
represented by entropy, increases. Therefore, one expects to see
the modification in temperature. But, the correction also affects
the internal energy on a certain level. This modification combined
with one in entropy provides a framework in which increasing the
disorder in the system will be defined by the modification in
internal energy on a level which results into the absence of
modification in temperature. Specifically speaking, modification
in entropy which is caused due to the presence of the first-order
correction is controlled by changes in internal energy which leads
to the temperature being independent of this parameter. This is
somehow similar to the isothermal process in which the temperature
remains the same. This is one of the most important properties of
black holes with the first-order correction. They present
isothermal-like behavior for modifications in correction
parameter.

The high energy limit of temperature is governed by the term with electric
charge while for the asymptotical behavior the cosmological term plays the
key role (see Eq. (\ref{tem})). It is a matter of calculation to obtain the
roots of the temperature in the following manner:%
\begin{equation}
r_{+}\left( T=0\right) =\pm \frac{\sqrt{2g\left( \varepsilon \right) \Lambda
\left( \varepsilon \right) \left[ kg\left( \varepsilon \right) +\sqrt{%
k^{2}g^{2}\left( \varepsilon \right) \pm 4q^{2}\left( \varepsilon \right)
f^{2}\left( \varepsilon \right) \Lambda (\varepsilon )}\right] }}{2\Lambda
\left( \varepsilon \right) }.  \label{rootT}
\end{equation}
Evidently, the existence of root of the solutions depends on satisfaction of
the following conditions:
\begin{equation*}
k^{2}\geq \frac{4q^{2}\left( \varepsilon \right) f^{2}\left( \varepsilon
\right) \Lambda (\varepsilon )}{g^{2}\left( \varepsilon \right) }%
,~~~\&~~~\Lambda (\varepsilon )\left[ kg\left( \varepsilon \right) \pm \sqrt{%
k^{2}g^{2}\left( \varepsilon \right) -4q^{2}\left( \varepsilon \right)
f^{2}\left( \varepsilon \right) \Lambda (\varepsilon )}\right] \geq 0.
\end{equation*}

Here too, similar to the mass case, we employed numerical
evaluation to
understand the behavior of the temperature in more details (see Fig. \ref%
{FigCT}). For dS case, depending on choices of different
parameters, the temperature could enjoy one of the following
cases: I) two roots with one maximum are located between them, II)
One maximum which is also the root of temperature and III) one
maximum without any root. In the first case, the physical
solutions (positive temperature) exists only between the roots,
whereas for the other two cases, the temperature is negative for
any horizon radius, therefore no physical black hole exists. As
for AdS case, temperature could have one of the following cases:
I) only one root and temperature is an increasing function of
horizon radius, II) one root with one extremum (extremum is
located after root) and III) one root, one maximum and one
minimum. Presence of extremum in temperature signals the existence
of phase transition point for the black holes. In other words,
extrema of the temperature are those points where the system goes
through phase transitions. Considering the behavior of the
temperature in dS and AdS cases, one can conclude that dS black
holes enjoy the existence of physical phase transition only in one
case (the case with two roots) whereas AdS ones, depending on
choices of different parameters, could have one or no phase
transition. Considering that thermal phase transition takes place
between thermally unstable black holes and stable ones, we need to
study the heat capacity to see what type of phase transitions are
present in the structure of these black holes for dS and AdS
cases.

The heat capacity, contrary to temperature is correction dependent.
Interestingly, the presence of correction parameter, $\alpha $, is only
observed in the numerator of heat capacity (\ref{heat}). This indicates that
the first-order correction only affects the places of root while the phase
transition points (divergencies of the heat capacity) are independent of
this parameter. It is a matter of calculation to obtain roots and
divergences of the heat capacity in the following form:
\begin{equation}
r_{+}\left( C=0\right) =\frac{1}{2^{4/3}}\sqrt{\frac{g\left( \varepsilon
\right) }{6\pi \Lambda \left( \varepsilon \right) }}\left( \frac{\mathcal{Y}%
}{\mathcal{P}}\right) ,  \label{rootC}
\end{equation}

\begin{equation}
r_{+}\left( C\rightarrow \infty \right) =\pm \frac{\sqrt{-2\Lambda \left(
\varepsilon \right) g\left( \varepsilon \right) \left[ kg\left( \varepsilon
\right) \pm \sqrt{k^{2}g^{2}\left( \varepsilon \right) +12q^{2}\left(
\varepsilon \right) f^{2}\left( \varepsilon \right) \Lambda (\varepsilon )}%
\right] }}{2\Lambda \left( \varepsilon \right) },  \label{divC}
\end{equation}%
in which
\begin{eqnarray*}
\mathcal{Y} &=&2^{4/3}\mathcal{Z}^{2/3}+2^{5/3}\pi kg\left( \varepsilon
\right) \mathcal{Z}^{1/3}+2^{14/3}g\left( \varepsilon \right) \sqrt{\alpha
\Lambda \left( \varepsilon \right) \mathcal{Z}^{1/3}} \\
&+&4g^{2}(\varepsilon )\left[ \pi k+4\alpha \Lambda \left( \varepsilon
\right) \right] ^{2}-12\pi ^{2}q^{2}(\varepsilon )f^{2}(\varepsilon )\Lambda
\left( \varepsilon \right) , \\
&& \\
\mathcal{P} &=&\sqrt{\pi \Lambda \left( \varepsilon \right) }\mathcal{Z}%
^{1/6}, \\
&& \\
\mathcal{Z} &=&2g(\varepsilon )\left[ g^{2}(\varepsilon )\left( \pi
k+4\alpha \Lambda \left( \varepsilon \right) \right) ^{3}-\frac{9\pi ^{2}}{2}%
q^{2}(\varepsilon )f^{2}(\varepsilon )\Lambda \left( \varepsilon \right) %
\left[ \pi k+16\alpha \Lambda \left( \varepsilon \right) \right] \right] \\
&+&3q(\varepsilon )f(\varepsilon )\Lambda \left( \varepsilon \right)
\mathcal{H}, \\
\mbox{with} && \\
\mathcal{H} &=&\left[ -48\alpha g^{4}(\varepsilon )\left[ \pi k+4\alpha
\Lambda \left( \varepsilon \right) \right] ^{3}-3\pi ^{2}q^{2}(\varepsilon
)g^{2}(\varepsilon )f^{2}(\varepsilon )\left[ \pi ^{2}k^{2}\right. \right. \\
&-&\left. \left. 704\alpha ^{2}\Lambda ^{2}\left( \varepsilon \right) -64\pi
\alpha k\Lambda \left( \varepsilon \right) \right] +12\pi
^{4}q^{4}(\varepsilon )f^{4}(\varepsilon )\Lambda \left( \varepsilon \right) %
\right] .
\end{eqnarray*}

Existence of the divergence point for the solutions is limited to
satisfaction of the following conditions
\begin{equation*}
k^{2}\geq -\frac{12\Lambda (\varepsilon )q^{2}\left( \varepsilon \right)
f^{2}\left( \varepsilon \right) }{g^{2}\left( \varepsilon \right) }%
,~~~\&~~~\Lambda (\varepsilon )\left[ -kg\left( \varepsilon \right) \pm
\sqrt{k^{2}g^{2}\left( \varepsilon \right) +12\Lambda (\varepsilon
)q^{2}\left( \varepsilon \right) f^{2}\left( \varepsilon \right) }\right]
\geq 0.
\end{equation*}

\begin{figure}[tbp]
$%
\begin{array}{cc}
\epsfxsize=7cm \epsffile{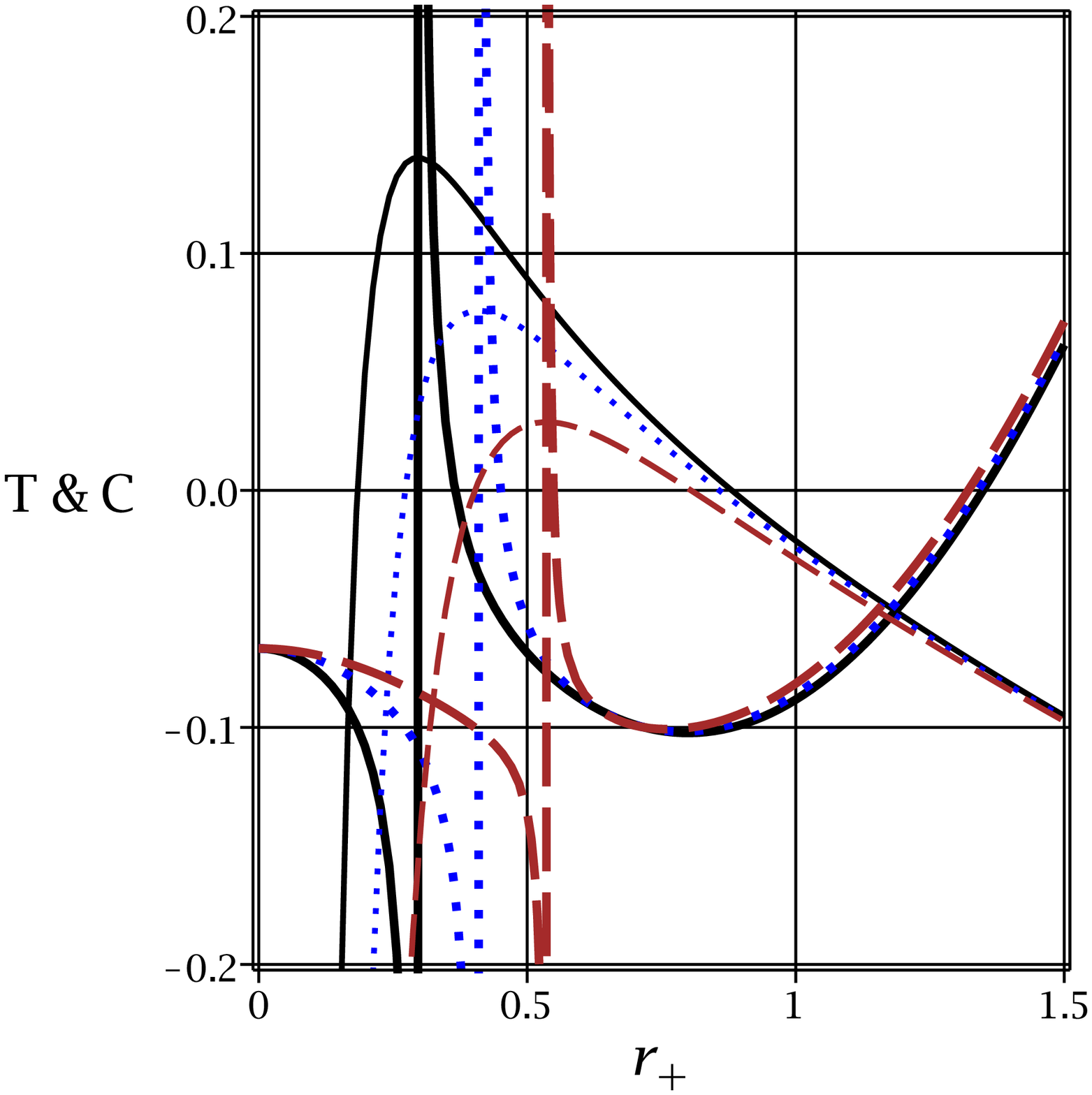} & \epsfxsize=7cm \epsffile{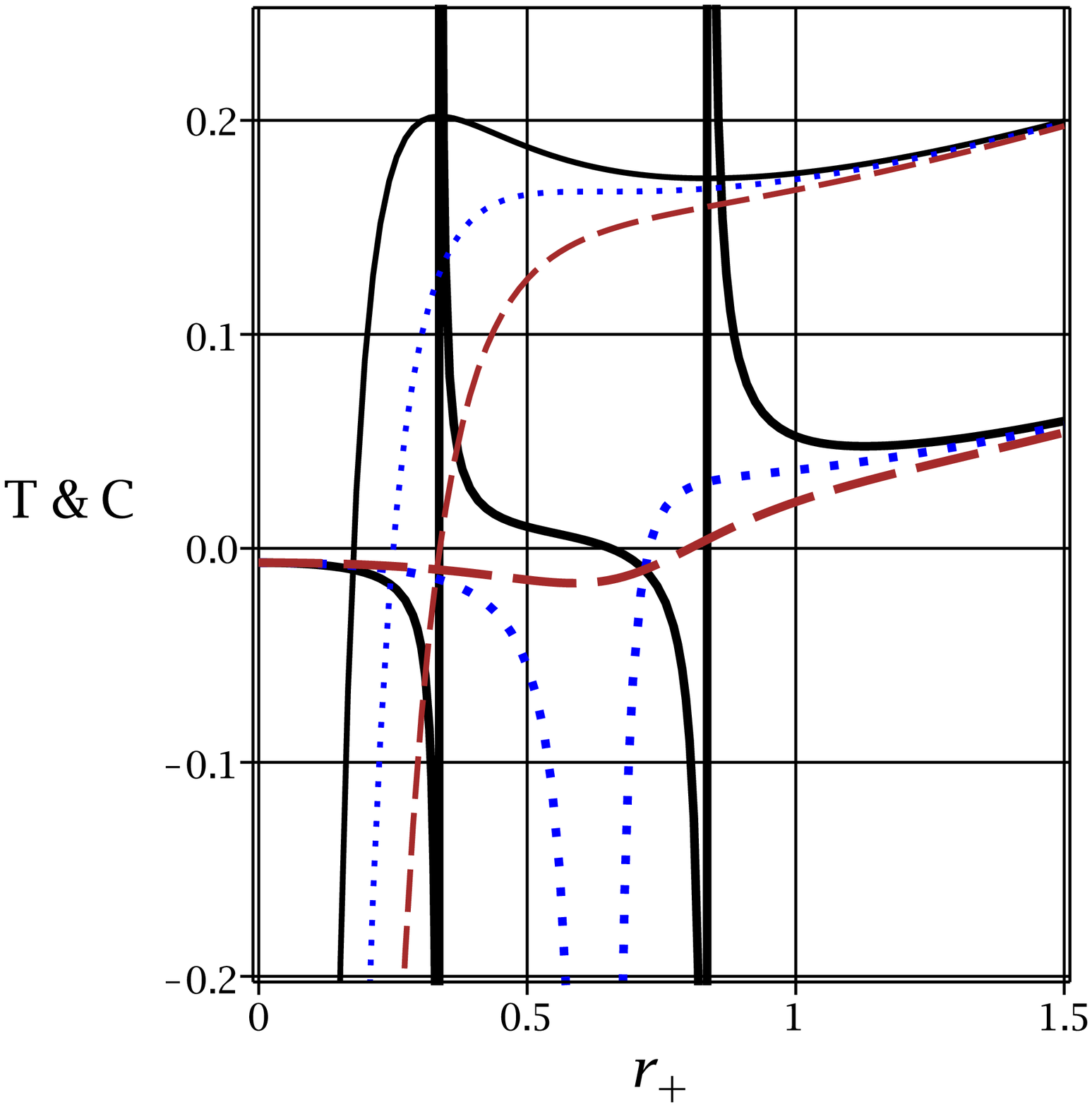}%
\end{array}
$%
\caption{$C$ (bold lines) and $T$ versus $r_{+}$ for $f(E)=g(\protect%
\varepsilon )=0.9$, $k=\protect\alpha =1$, $q(\protect\varepsilon )=0.2$
(continuous line), $q(\protect\varepsilon )=0.2887$ (dotted line) and $q(%
\protect\varepsilon )=0.4$ (dashed line). \newline
Left panel: $\Lambda (\protect\varepsilon )=1$; Right panel: $\Lambda (%
\protect\varepsilon )=-1$.}
\label{FigCT}
\end{figure}

The first condition is always satisfied for dS black holes.
Considering that square root function is always positive valued
and by taking the structure of conditions into account, one can
conclude that there is always a divergence point for the dS black
holes. In addition, the dS case enjoys the existence of two roots
in its heat capacity. As for the AdS black holes, the situation is
more complicated. Depending on choices of different parameters,
these black holes could enjoy one of the following cases: I) two
divergence points with one root located between them, II) one
divergence point and one root in which root is observed after
divergence point, and III) only one root is observable. In order
to elaborate these results, we have plotted the diagrams (see Fig.
\ref{FigCT}).

\begin{figure}[tbp]
$%
\begin{array}{cc}
\epsfxsize=7cm \epsffile{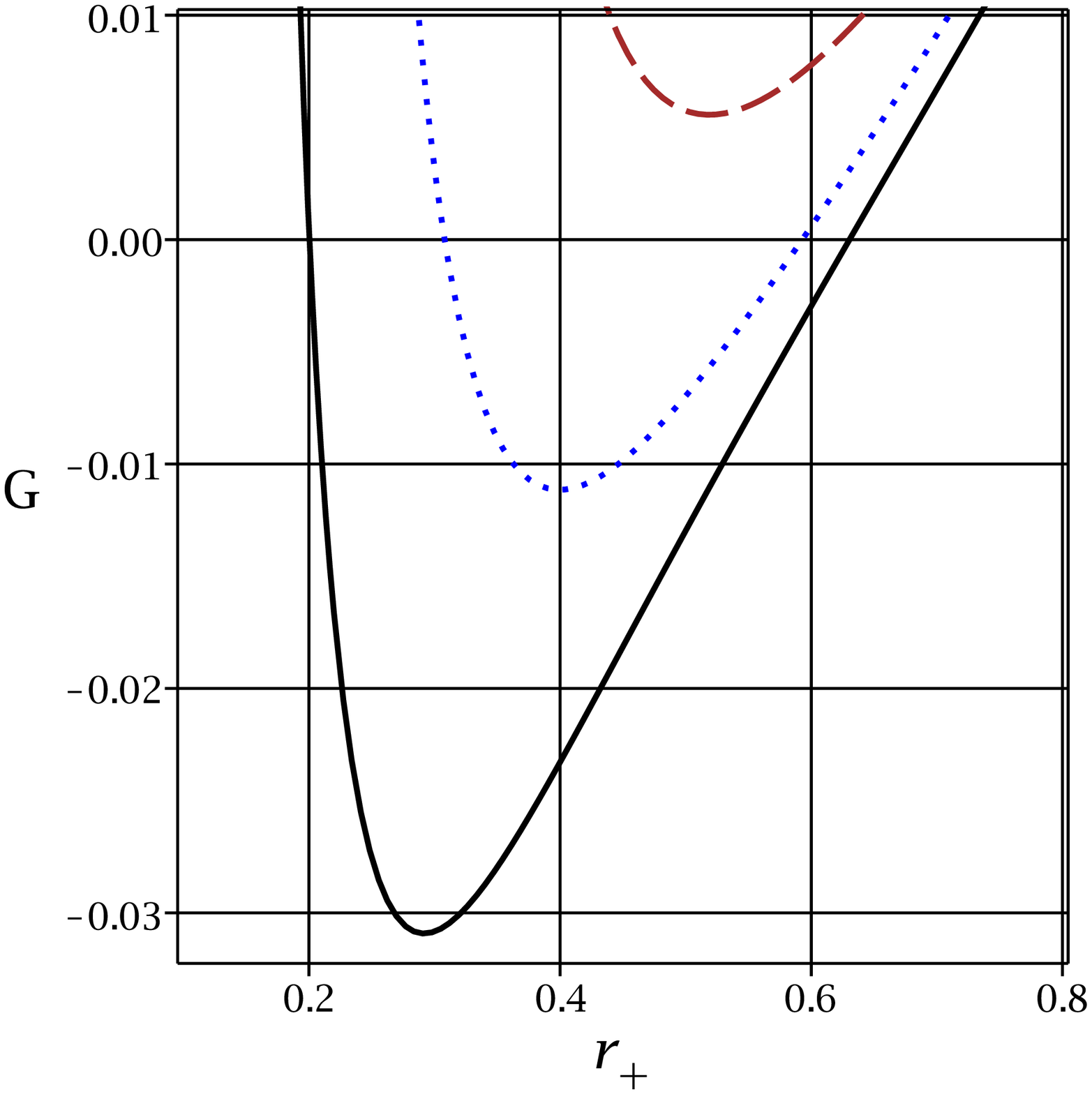} & \epsfxsize=7cm \epsffile{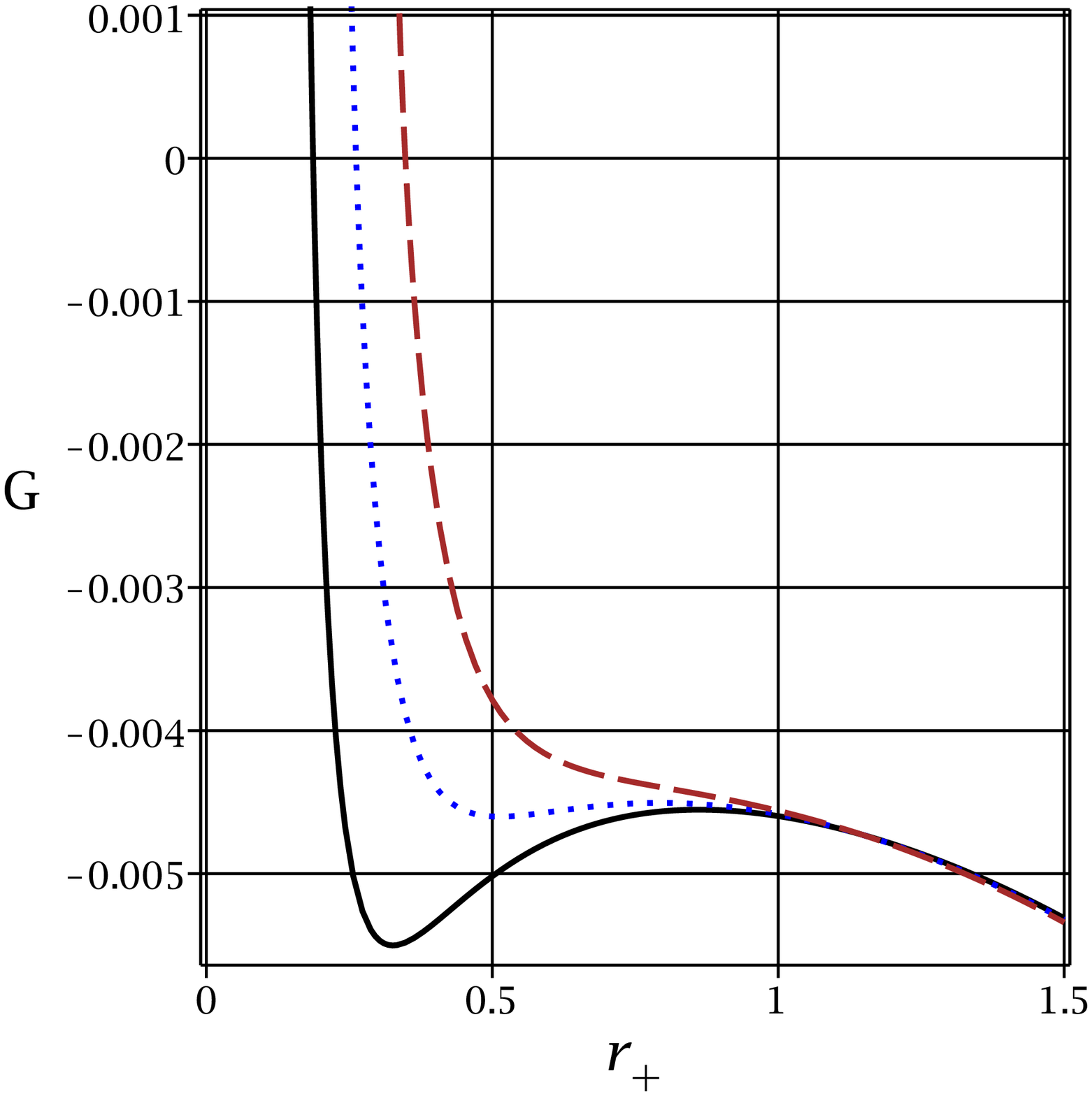}%
\end{array}
$%
\caption{$G$ versus $r_{+}$ for $f(\protect\varepsilon )=g(\protect%
\varepsilon )=0.9$, $k=\protect\alpha =1$, $q(\protect\varepsilon )=0.2$
(continuous line), $q(\protect\varepsilon )=0.2887$ (dotted line) and $q(%
\protect\varepsilon )=0.4$ (dashed line). \newline
Left panel: $\Lambda (\protect\varepsilon )=1$; Right panel: $\Lambda (%
\protect\varepsilon )=-1$.}
\label{FigG}
\end{figure}


\begin{figure}[tbp]
$%
\begin{array}{cc}
\epsfxsize=7cm \epsffile{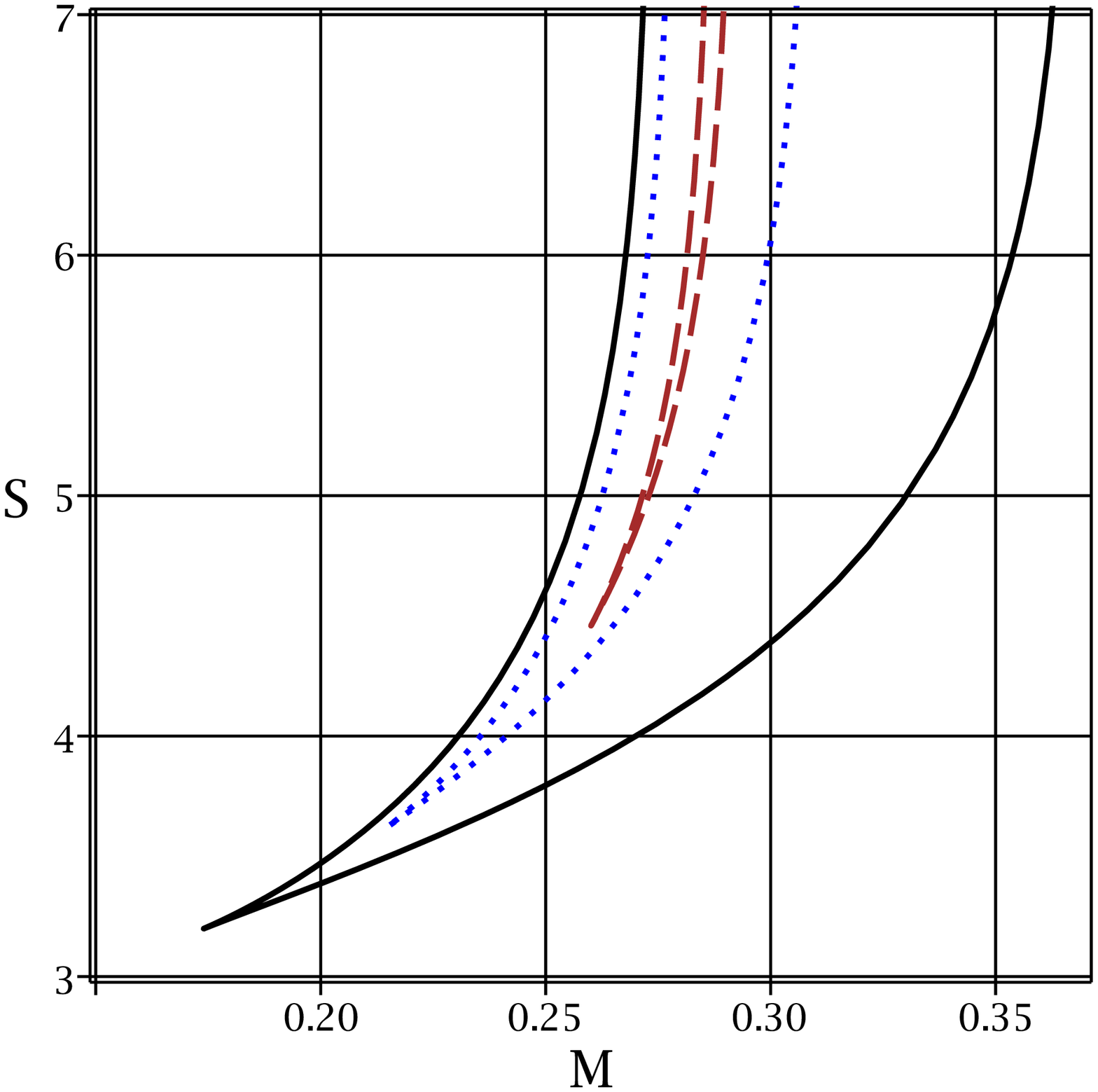} & \epsfxsize=7cm \epsffile{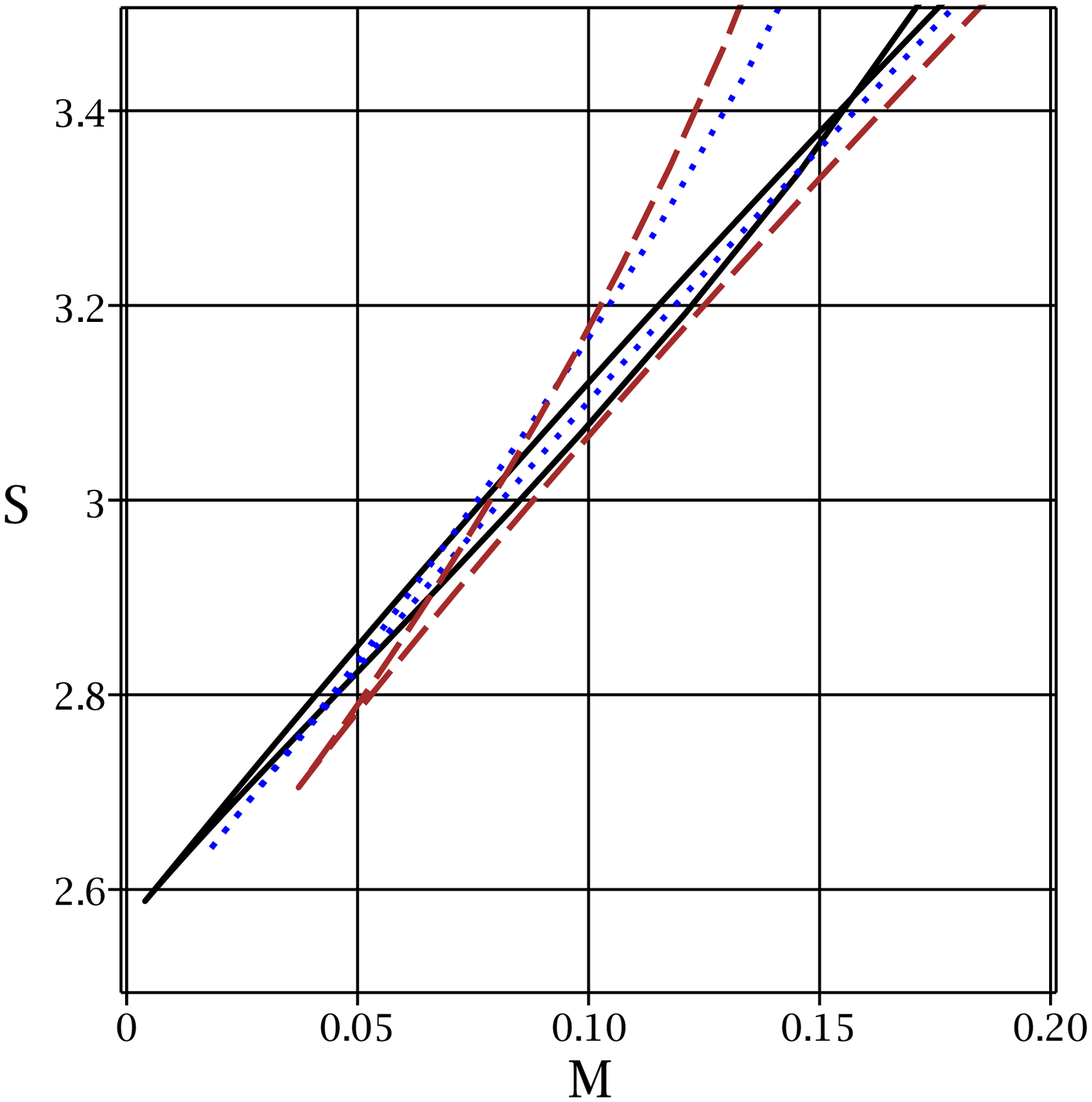}%
\end{array}
$%
\caption{$S$ versus $M$ for $f(\protect\varepsilon )=g(\protect\varepsilon %
)=0.9$, $k=\protect\alpha =1$, $q(\protect\varepsilon )=0.2$ (continuous
line), $q(\protect\varepsilon )=0.2887$ (dotted line) and $q(\protect%
\varepsilon )=0.4$ (dashed line). \newline
Left panel: $\Lambda (\protect\varepsilon )=1$; Right panel: $\Lambda (%
\protect\varepsilon )=-1$.}
\label{FigSM}
\end{figure}

\begin{figure}[tbp]
$%
\begin{array}{cc}
\epsfxsize=7cm \epsffile{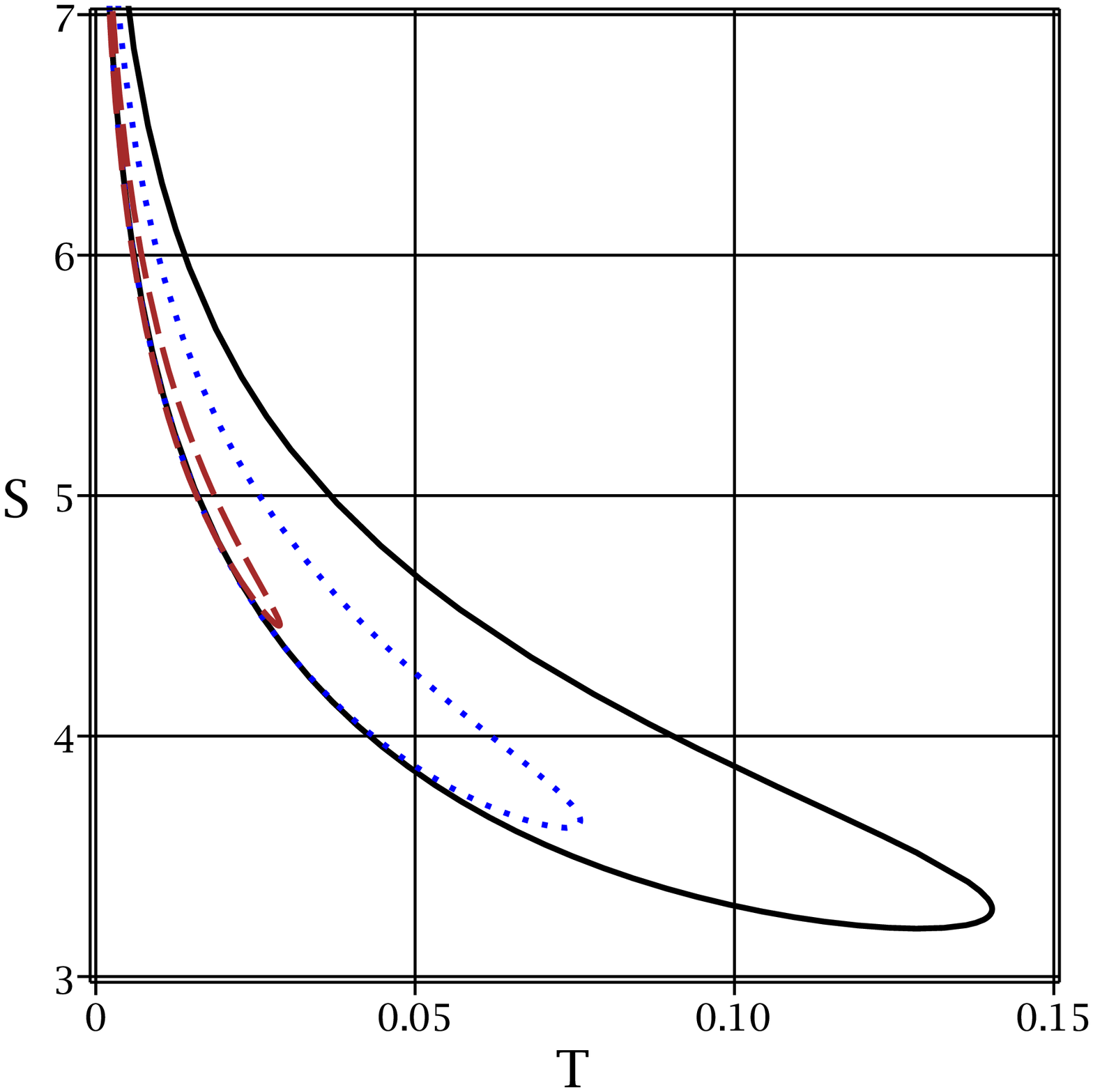} & \epsfxsize=7cm \epsffile{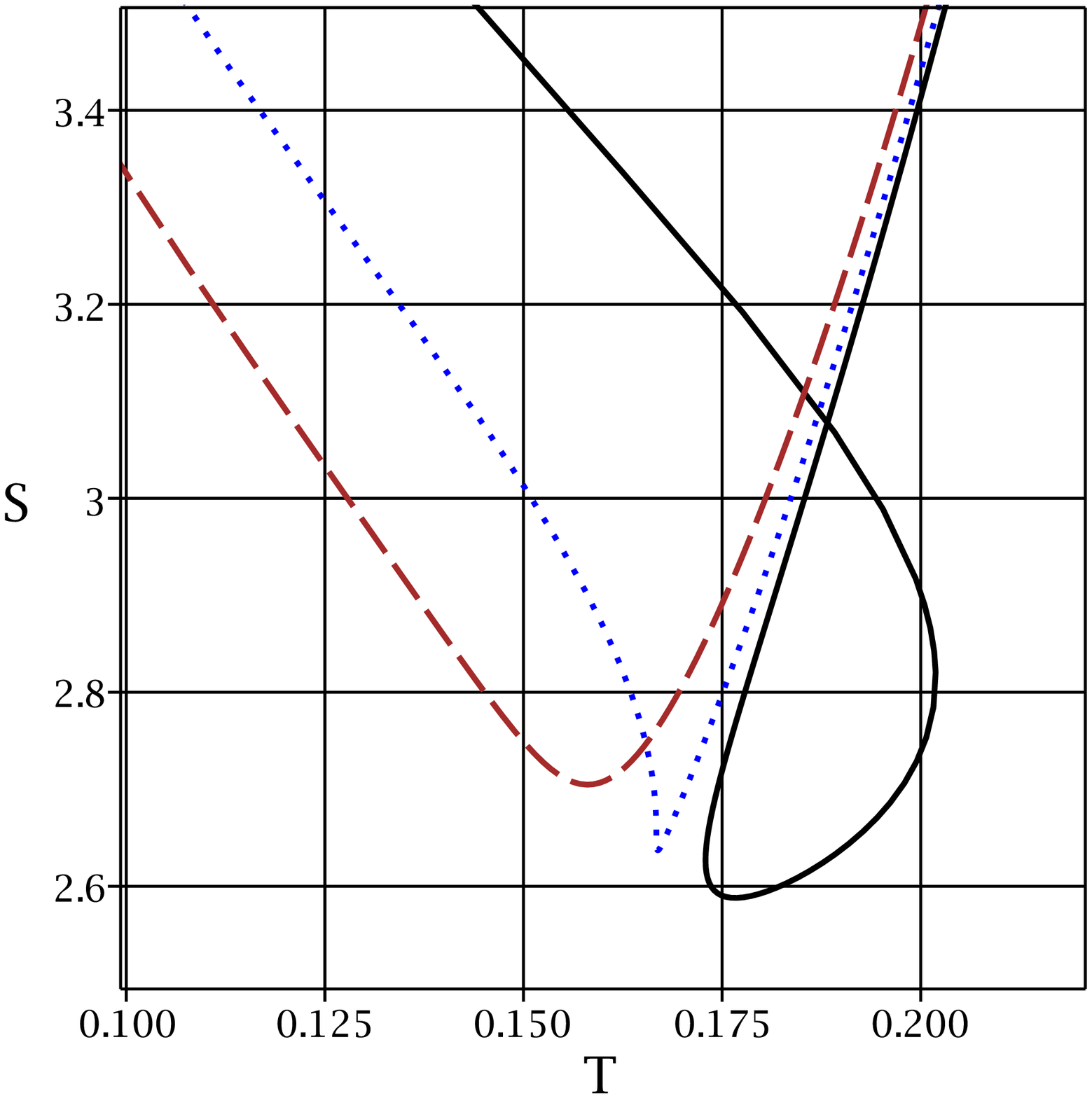}%
\end{array}
$%
\caption{$S$ versus $T$ for $f(\protect\varepsilon )=g(\protect\varepsilon %
)=0.9$, $k=\protect\alpha =1$, $q(\protect\varepsilon )=0.2$ (continuous
line), $q(\protect\varepsilon )=0.2887$ (dotted line) and $q(\protect%
\varepsilon )=0.4$ (dashed line). \newline
Left panel: $\Lambda (\protect\varepsilon )=1$; Right panel: $\Lambda (%
\protect\varepsilon )=-1$.}
\label{FigST}
\end{figure}


In order to make accurate predictions regarding thermodynamical
properties of these black holes, one must consider the
thermodynamical behavior of temperature alongside with the heat
capacity. The reason is that negativity/positivity of the
temperature enforces s harder restriction on values that different
parameters can acquire. This is due to fact that negative values
of the temperature are considered as non-physical cases. For the
dS case, the heat capacity enjoys the existence of one divergence
and two roots. The roots are located after the divergence. Between
the root of the temperature and divergence of the heat capacity,
the heat capacity is negative valued and solutions are thermally
unstable. At the divergence of heat capacity, a phase transition
between smaller unstable and larger stable black holes takes
place. Between the divergence point and the smaller root of the
heat capacity, solutions are thermally stable. Whereas, between
the smaller root of heat capacity and the larger root of the
temperature, solutions are thermally unstable. The larger root of
heat capacity is located after the larger root of temperature.
Although the heat capacity is positive after its larger root,
since the temperature is negative here, no physical black hole
exists in this region.

As for the AdS case, thermal stability depends on the number of
divergences and the place of root. In the absence of divergence,
solutions are thermally unstable between the root of temperature
and the root of heat capacity. Whereas, after the root of heat
capacity, solutions are physical and stable. In case of the
existence of one divergence for heat capacity, interestingly,
between the root of temperature and divergence of the heat
capacity, and also between the divergence point of heat capacity
and its root, the solutions are thermally unstable. The only
stable region is after the root of heat capacity. This indicates
that although there is phase transition for AdS black holes, the
phase transition occurs between two unstable black holes.
Thermally stable black holes are obtained only after the root of
heat capacity. Finally, in case of two divergences, the only
positive regions for heat capacity are between the smaller
divergence of heat capacity and its root, and after larger
divergence. Therefore, there are two phase transitions: one
between smaller unstable and medium stable black holes; the other
is between medium unstable and larger stable black holes. This is
due to the fact that in the root of heat capacity which is located
between the divergences, there is a change of signature taking
place.

One of the most interesting effects of the first-order correction could be
detected through the high energy limit of heat capacity. It is a matter of
calculation to show that high energy of this quantity is
\begin{equation*}
\lim_{r_{+}\rightarrow 0}C=-\frac{2}{3}\alpha -\frac{\pi q^{2}\left(
\varepsilon \right) f^{2}\left( \varepsilon \right) +\frac{4}{3}\alpha
kg^{2}\left( \varepsilon \right) }{6q^{2}\left( \varepsilon \right)
g^{2}\left( \varepsilon \right) f^{2}\left( \varepsilon \right) }r_{+}^{2}+%
\mathcal{O}\left( r_{+}^{4}\right) .
\end{equation*}

Here, we see that high energy limit of the heat capacity is purely
governed by correction parameter, $\alpha $, without any factor of
other quantities. Considering the positive $\alpha $, one can find
that the heat capacity is negative in the high energy limit. In
addition, since the correction factor is not coupled with any
other quantity, it indicates that if the black hole vanishes,
there is a non-zero heat capacity available which could mark the
existence of the black hole. This means that all the information
regarding the existence of the black hole is not lost and some of
it remains in form of non-vanishing heat capacity. This is another
important property of the first-order correction which shows the
deviation from usual black holes thermodynamics. On the other
hand, the asymptotical behavior of the heat capacity is given by
\begin{eqnarray}
\lim_{r_{+}\rightarrow \infty }C &=&\frac{\pi r_{+}^{2}}{2g^{2}\left(
\varepsilon \right) }-\frac{\pi k+2\alpha \Lambda \left( \varepsilon \right)
}{\Lambda \left( \varepsilon \right) }+\frac{2\pi q^{2}\left( \varepsilon
\right) f^{2}\left( \varepsilon \right) \Lambda \left( \varepsilon \right)
+kg^{2}\left( \varepsilon \right) \left( \pi k+2\alpha \Lambda \left(
\varepsilon \right) \right) }{2\Lambda ^{2}\left( \varepsilon \right)
r_{+}^{2}}  \notag \\
&+&\mathcal{O}\left( \frac{1}{r_{+}^{4}}\right) .
\end{eqnarray}

Evidently, the asymptotical behavior of the solutions is governed
only by gravitational part. This is due to fact that in the
dominant term, no traces of the electric charge, the first-order
correction, topological factor and cosmological constant are
observable. Therefore, the asymptotical behavior is governed by
gravitational part purely. It is worthwhile to mention that the
effects of the first-order correction could be observed in the
second and third dominant terms.

Next, we study the behavior of Gibbs free energy. According to the
thermodynamical principle, the Gibbs free energy could be employed
to extract the phase transition points. Phase transition points in
Gibbs free energy can be present at extrema. Therefore, the
existence of extrema in Gibbs free energy diagrams indicates that
solutions enjoy thermal phase transitions in their structure.
Unfortunately, due to the complexity of the obtained Gibbs free
energy, it was not possible to obtain the extrema of it
analytically. Therefore, we employed the numerical method (see
Fig. \ref{FigG}). Evidently, for the dS case, Gibbs free energy
enjoys the existence of a minimum in its structure. The place of
this minimum is exactly the same as divergence of the heat
capacity. Therefore, the phase transition point observed in the
heat capacity coincides with the one in the Gibbs free energy. It
is worthwhile to mention that Gibbs free energy, in this case,
could enjoy one of the following cases as well: I) absence of
root, II) the existence of one extreme root, and III) presences of
two roots. The AdS case presents the larger variation in its
behavior. It could have: I) only one root, II) one root with one
extremum, and III) one root with two extrema. The extrema in any
of the mentioned cases are located exactly where the heat capacity
meets divergences.

As for the high energy limit, interestingly, similar to the heat capacity
case, the high energy limit is governed by correction factor, $\alpha $,
\begin{eqnarray}
\lim_{r_{+}\rightarrow 0}G &=&-\frac{\alpha q^{2}\left( \varepsilon \right)
g\left( \varepsilon \right) f\left( \varepsilon \right) \left[ 3\ln \left(
\frac{-q^{2}\left( \varepsilon \right) f\left( \varepsilon \right) }{8\sqrt{%
\pi }r_{+}^{2}}\right) -2\right] }{12\pi r_{+}^{3}}  \notag \\
&&-\frac{4\alpha kg^{2}\left( \varepsilon \right) -\pi q^{2}\left(
\varepsilon \right) f^{2}\left( \varepsilon \right) +4\alpha kg^{2}\left(
\varepsilon \right) \ln \left( \frac{-q^{2}\left( \varepsilon \right)
f\left( \varepsilon \right) }{8\sqrt{\pi }r_{+}^{2}}\right) }{16\pi g\left(
\varepsilon \right) f\left( \varepsilon \right) r_{+}}+\mathcal{O}\left(
r_{+}\right) ,
\end{eqnarray}%
but contrary to heat capacity case, here, there is a coupling between
correction term and electric charge part of the solutions. On the other
hand, the dominant term in asymptotical behavior of the Gibbs free energy is
the cosmological constant term without any coupling with correction term
which can be seen from following:
\begin{eqnarray}
\lim_{r_{+}\rightarrow \infty }G &=&\frac{\Lambda \left( \varepsilon \right)
r_{+}^{3}}{48g^{3}\left( \varepsilon \right) f\left( \varepsilon \right) }-%
\frac{\left[ 8\alpha \Lambda \left( \varepsilon \right) \ln \left(
r_{+}\right) +4\alpha \Lambda \left( \varepsilon \right) \ln \left( \frac{%
-\Lambda \left( \varepsilon \right) }{8\sqrt{\pi }g^{2}\left( \varepsilon
\right) f\left( \varepsilon \right) }\right) -\left( \pi k+8\alpha \Lambda
\left( \varepsilon \right) \right) \right] r_{+}}{16\pi g\left( \varepsilon
\right) f\left( \varepsilon \right) }  \notag \\
&&+\frac{\frac{\alpha kg\left( \varepsilon \right) }{4\pi f\left(
\varepsilon \right) }\left[ \ln \left( \frac{-\Lambda \left( \varepsilon
\right) }{8\sqrt{\pi }g^{2}\left( \varepsilon \right) f\left( \varepsilon
\right) }\right) +1+2\ln \left( r_{+}\right) \right] -\frac{q^{2}\left(
\varepsilon \right) f\left( \varepsilon \right) }{16g\left( \varepsilon
\right) }}{r_{+}}+\mathcal{O}\left( \frac{1}{r_{+}^{3}}\right) .
\end{eqnarray}
In general, one can state that the effects of the first-order
correction term could be highly detected for small and medium
black holes while for the large black holes, it is not on
significant level.

Our next subject of the interest is phase diagrams for corrected entropy
versus effective mass (Fig. \ref{FigSM}) and temperature (Fig. \ref{FigST}).

For the case of entropy versus mass, evidently, irrespective of
the background spacetime (being dS or AdS), there exists a minimum
in which both the entropy and internal energy are minimum. The
entropy for both of the cases of dS and AdS is an increasing
function of the internal energy. In dS case, interestingly, two
branches exist in which for every entropy, there exists at least
two different internal energy. The only exception is the extremum
point (minimum) in which the two branches coincided. As for the
AdS case, the situation is same except for one of the diagrams. In
this case, two points exist in which two branches coincide. One of
them is the minimum that was discussed for the dS case, while the
other one is after the minimum. Interestingly, at this point (see
the continuous line in the right panel of Fig. \ref{FigSM}), the
place of branches is switched. Let's say that before this point,
for any entropy, two internal energy exists: one belongs to branch
$A$ (right branch in Fig. \ref{FigSM}) and the other one belongs
to branch $B$ (right branch in Fig. \ref{FigSM}) in which internal
energy of branch $B$ is bigger than the one in branch $A$. After
the mentioned point, the bigger internal energy will belong to
branch $A$ and the smaller one will be of branch $B$.

The situation in entropy versus temperature diagrams is different
for dS and AdS cases. Here, for the dS case, the temperature has a
maximum that can be acquired. The maximum is actually where
entropy acquires its minimum. Here too, for every entropy, two
temperature exist except for the minimum where both branches
coincide. On the other hand, the AdS case situation is different.
Here, there is no maximum present for the temperature while one
can obtain a minimum for the entropy. Before the minimum, entropy
is a decreasing function of the temperature, while after the
minimum, the entropy will be an increasing function of the
temperature. Interestingly, for specific cases (see the continuous
line in the left panel of Fig. \ref{FigST}), it is possible for
the two branches of entropy versus temperature to coincide in two
points and form a closed cycle. In this case, the entropy is first
a decreasing function of the temperature, then both of the
temperature and entropy decreases to reach the minimum point.
After that, the entropy becomes an increasing function of the
temperature.

The plotted diagrams for the dS case have yet another property
which was absent in AdS diagram except for the mentioned special
case. This property is due to the fact that for every temperature
there are two entropies while for every entropy, there are also
two temperatures. In a manner, the plotted diagrams for the dS
case are closed one. For the AdS diagrams, closed diagram was
observed only for the special case where two coinciding points for
the two branches exist. A simple comparison between these diagrams
with those plotted before shows that this special case for AdS
background is related to the presence of two divergences in the
heat capacity diagrams. This confirms that entropy versus
temperature diagrams recognize the existence of phase transition
point in its structure. Here, we see that background spacetime has
profound contributions to the thermodynamical structure of the
black holes. Plotted diagrams for entropy versus temperature
measures the changes in entropy by temperature fluctuation.
Evidently, for the dS case, we come across a cyclic (closed)
diagrams indicating the existence of bounds on values that
temperature could acquire. On the other hand, for the AdS case,
such closed cyclic diagrams were observed for specific values of
different parameters but the existence of this cycle does not
bound the temperature. In fact, it puts a bound on limits that
entropy can acquire. Here, we see that the entropy could not
vanish completely.

\section{Correction versus non-correction}

\begin{figure}[tbp]
$%
\begin{array}{cc}
\epsfxsize=7cm \epsffile{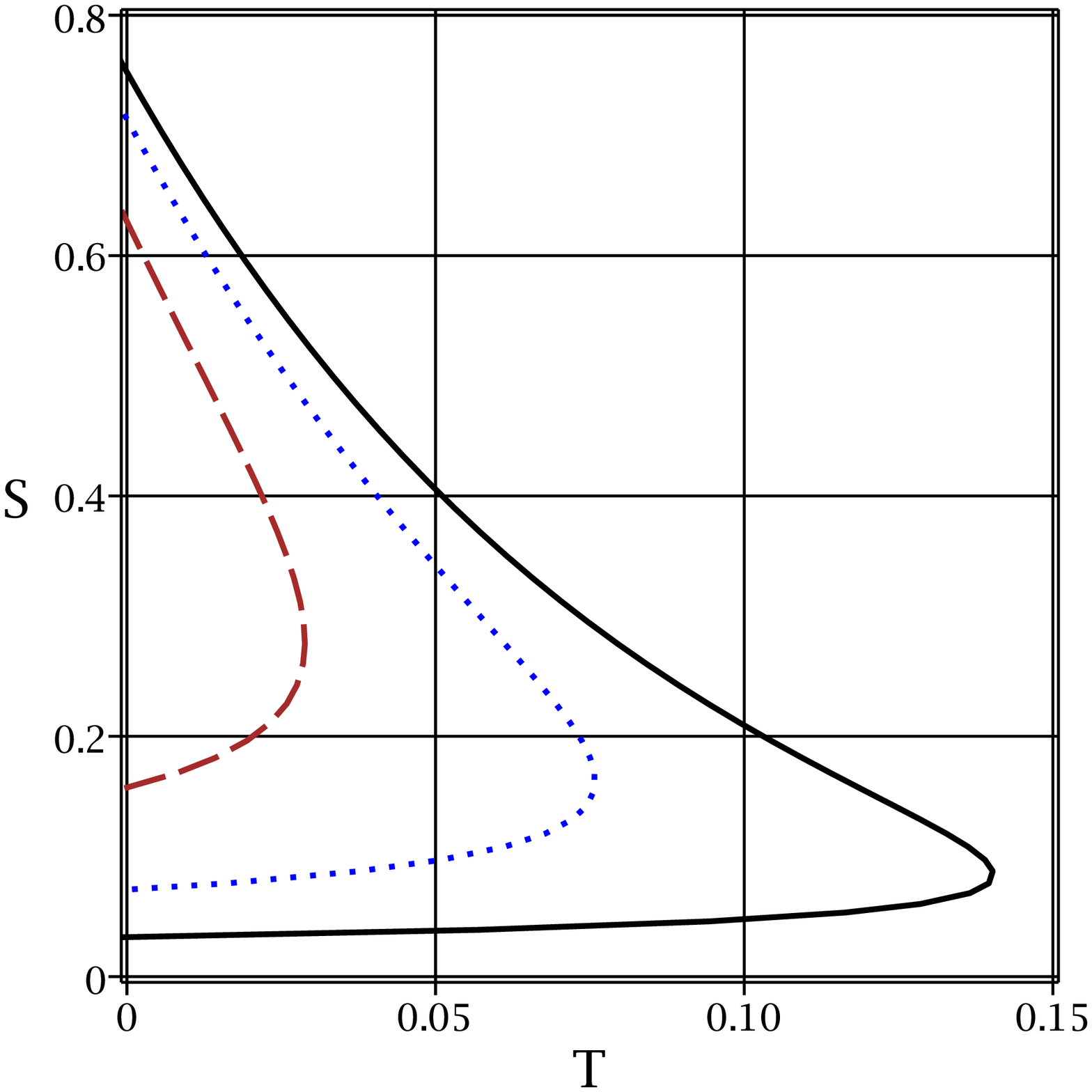} & \epsfxsize=7cm \epsffile{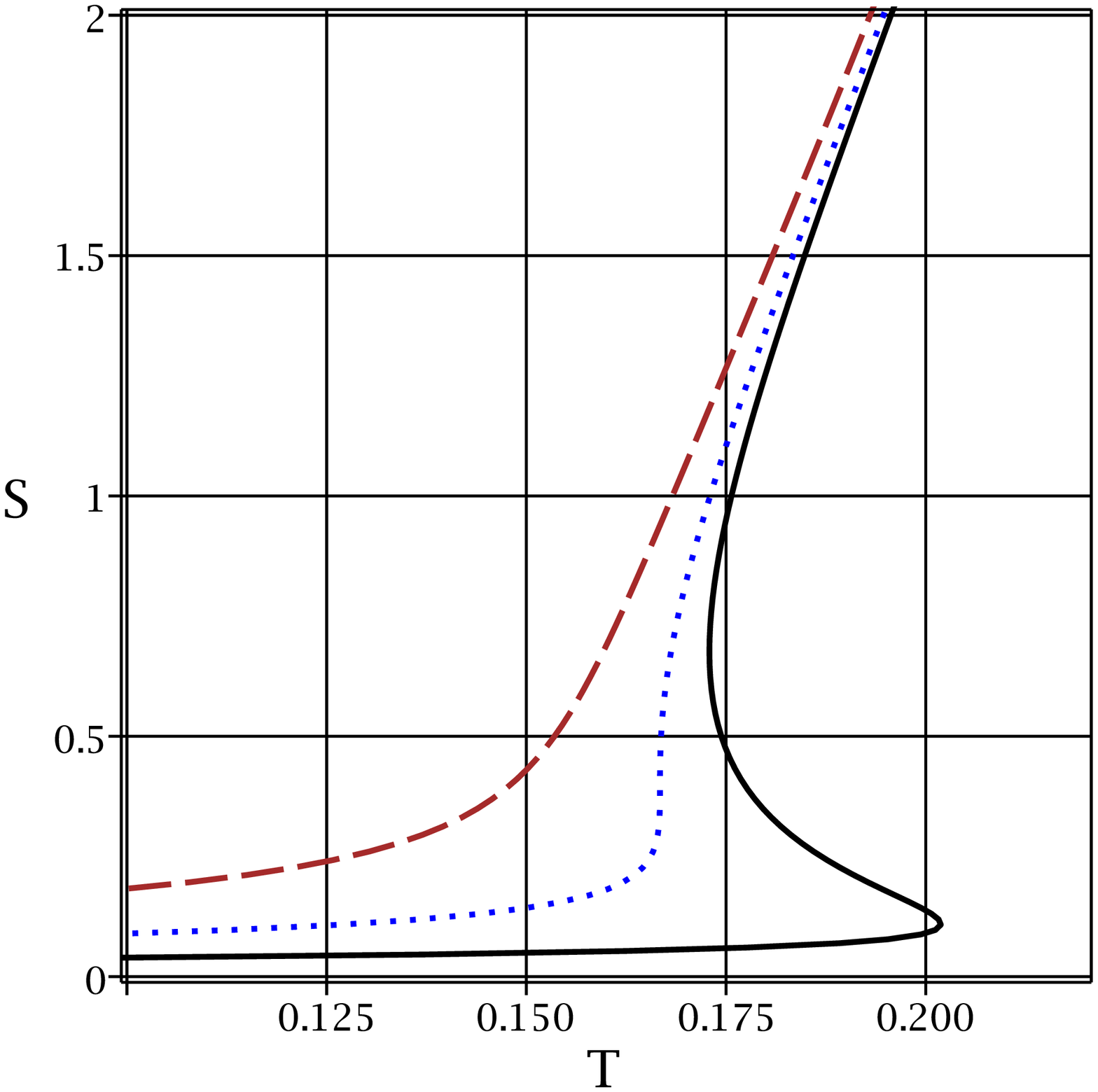}%
\end{array}
$%
\caption{$S$ versus $T$ for $\protect\alpha =0$, $f(\protect\varepsilon )=g(%
\protect\varepsilon )=0.9$, $k=1$, $q(\protect\varepsilon )=0.2$ (continuous
line), $q(\protect\varepsilon )=0.2887$ (dotted line) and $q(\protect%
\varepsilon )=0.4$ (dashed line). \newline
Left panel: $\Lambda (\protect\varepsilon )=1$; Right panel: $\Lambda (%
\protect\varepsilon )=-1$.}
\label{FigSTT}
\end{figure}

The final section of this paper is devoted to understand the
effects of the first-order correction and compare it with the
non-correction case. Here, we intend to see how the entropy and
stability are modified due to contributions of the first-order
correction. First of all, we have plotted diagrams for the entropy
versus temperature in the absence of the first-order correction,
i.e. $\alpha=0$, in Fig. \ref{FigSTT}.

\begin{figure}[tbp]
$%
\begin{array}{cc}
\epsfxsize=7cm \epsffile{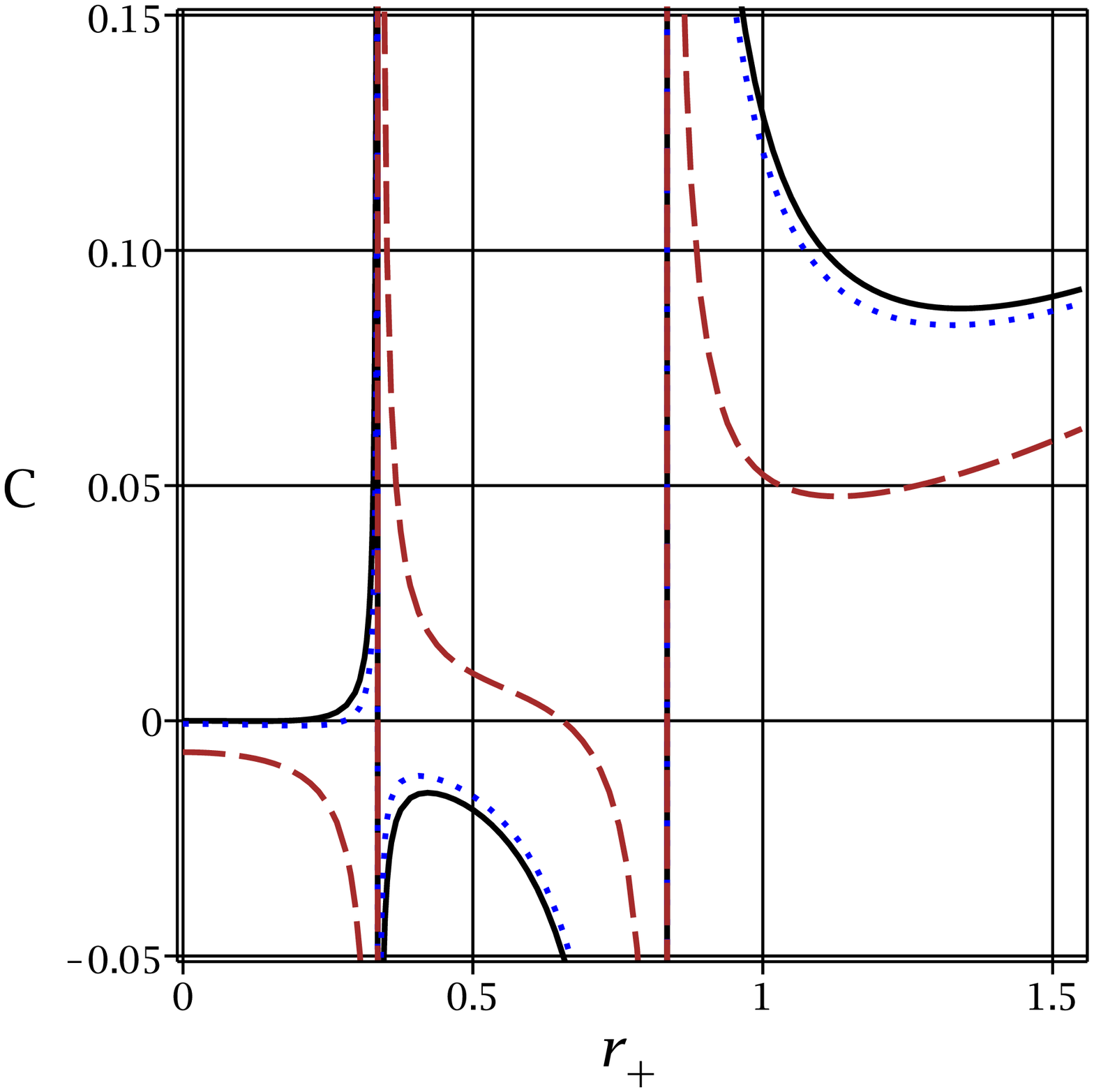} & \epsfxsize=7cm \epsffile{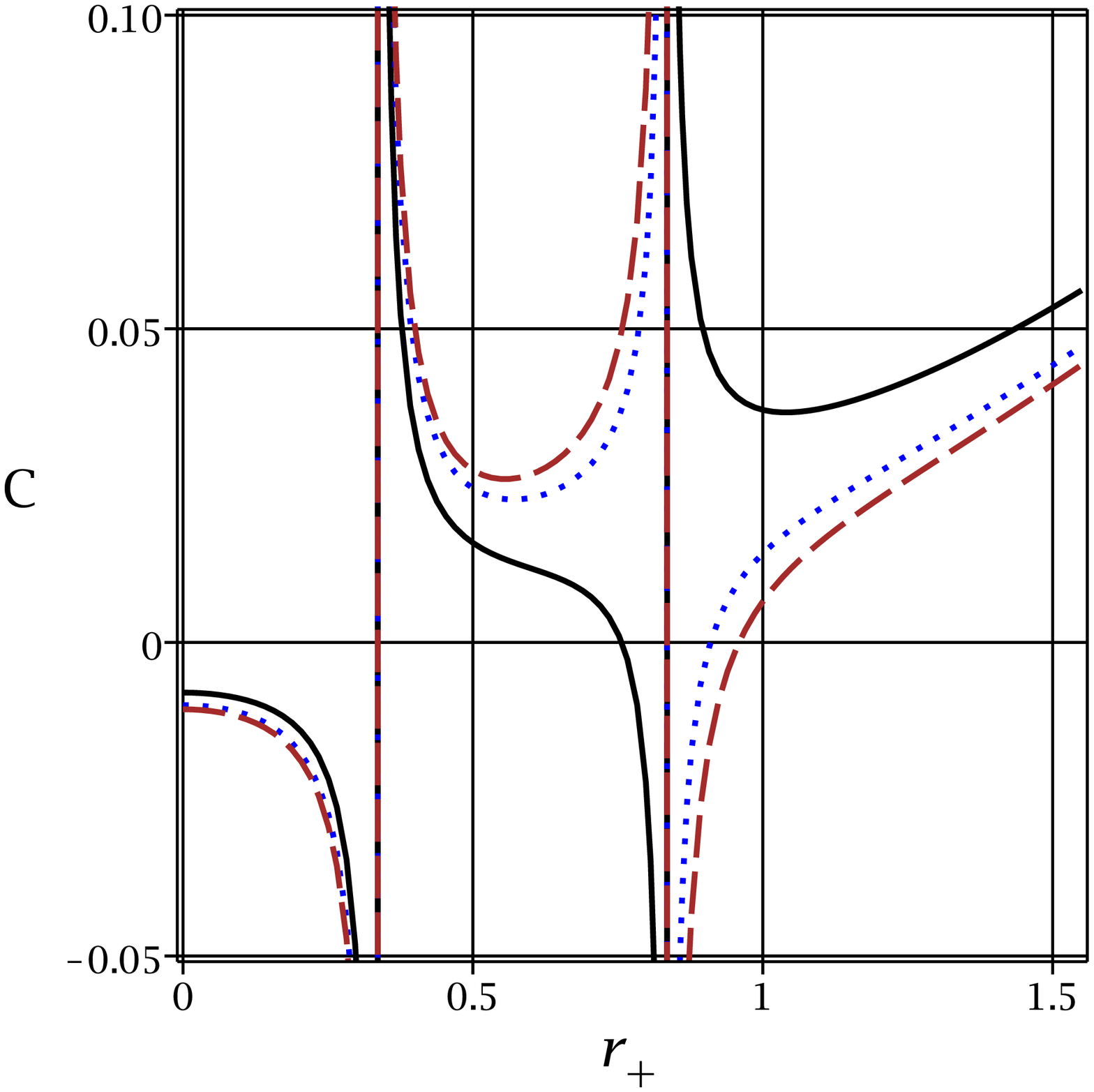}%
\end{array}
$%
\caption{$C$ versus $r_{+}$ for $\Lambda (\protect\varepsilon )=-1$, $f(%
\protect\varepsilon )=g(\protect\varepsilon )=0.9$, $k=1$ and $q(\protect%
\varepsilon )=0.2$. \newline
Left panel: $\protect\alpha =0$ (continuous line), $\protect\alpha =0.1$
(dotted line) and $\protect\alpha =1$ (dashed line). \newline
Right panel: $\protect\alpha =1.2$ (continuous line), $\protect\alpha =1.5$
(dotted line) and $\protect\alpha =1.6$ (dashed line).}
\label{FigCC1}
\end{figure}

\begin{figure}[tbp]
$%
\begin{array}{cc}
\epsfxsize=7cm \epsffile{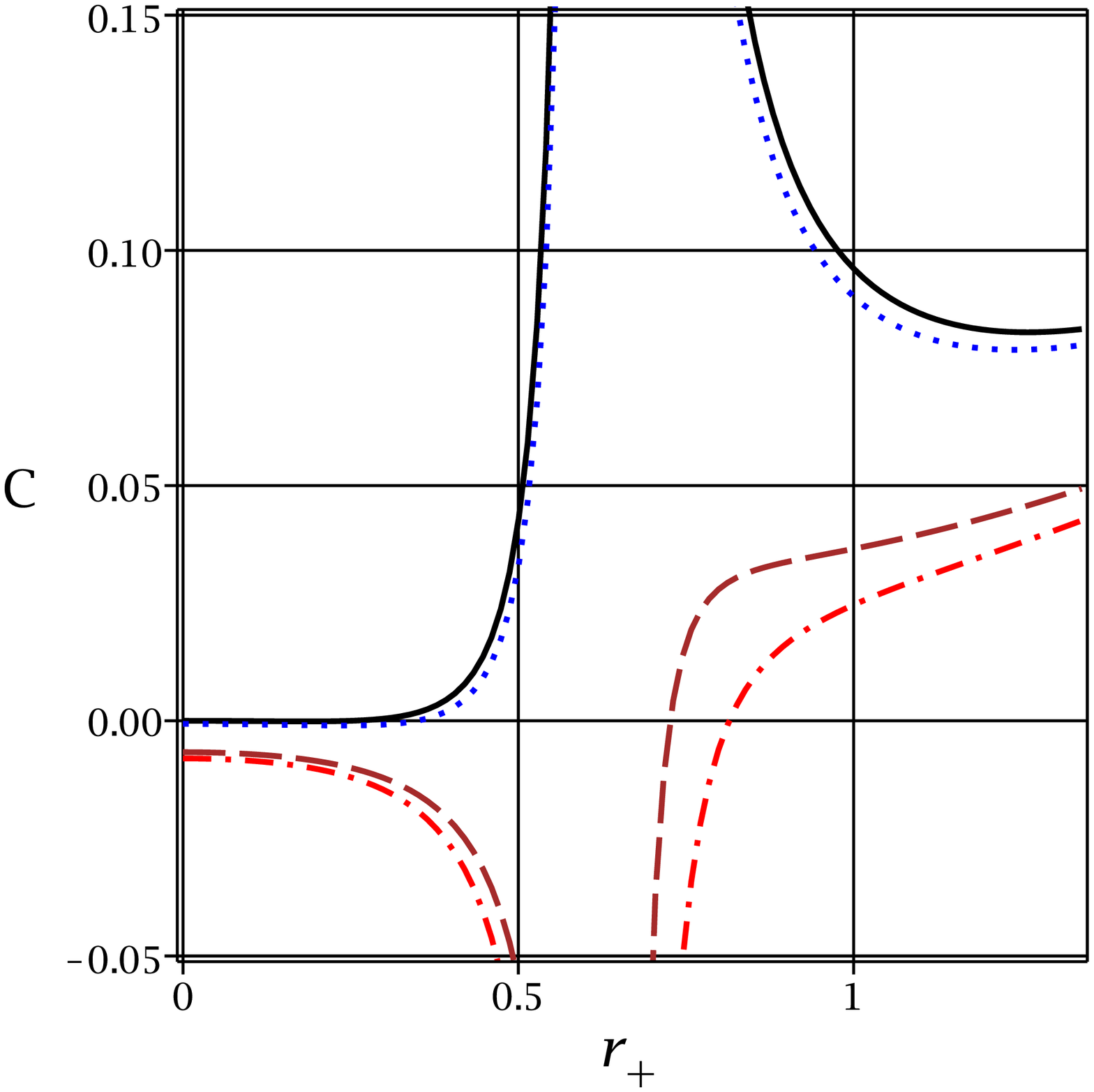} & \epsfxsize=7cm \epsffile{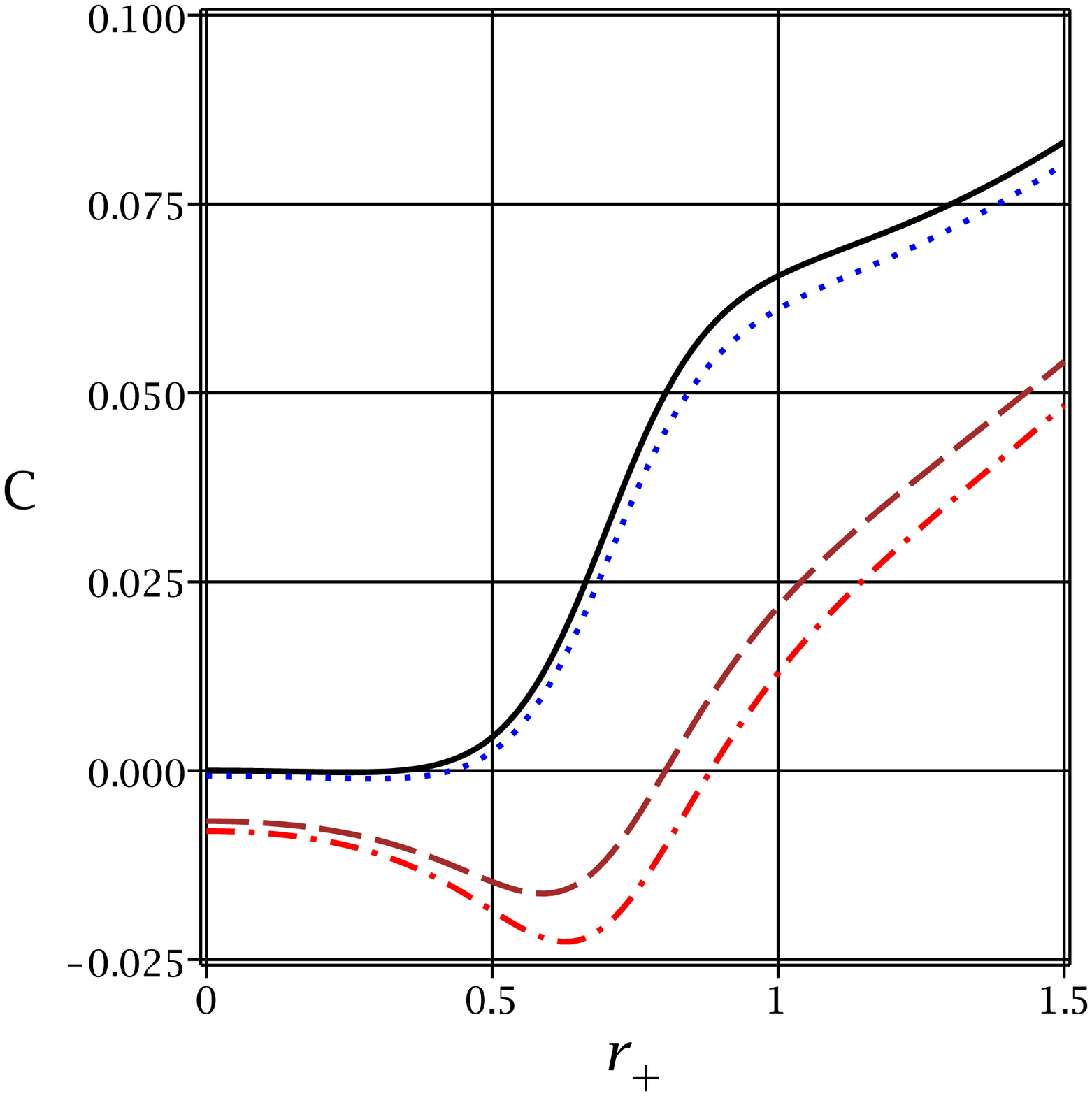}%
\end{array}
$%
\caption{$C$ versus $r_{+}$ for $\Lambda (\protect\varepsilon )=-1$, $f(%
\protect\varepsilon )=g(\protect\varepsilon )=0.9$, $k=1$, $\protect\alpha %
=0 $ (continuous line), $\protect\alpha =0.1$ (dotted line), $\protect\alpha %
=1$ (dashed line) and $\protect\alpha =1$ (dashed-dotted line). \newline
Left panel: $q(\protect\varepsilon )=0.2887$; Right panel: $q(\protect%
\varepsilon )=0.4$.}
\label{FigCC2}
\end{figure}


In the case of dS, interestingly, we observe that no cyclic like
behavior is present and for vanishing the temperature, there are
two positive values available for the entropy of the system. In
this case, similar to the presence of correction, there is a
maximum available for the temperature and a minimum for entropy
but contrary to correction case, the minimum of entropy and
maximum of the temperature do not take place at the same point. In
the correction case, for every entropy (temperatures), there exist
two temperatures (entropies). In the absence of correction, for
every temperature, there exist two entropies. The only exception
is at the maximum of temperature.

The AdS case provides us with more interesting details. The
dashed, dotted and continuous lines in the right panel of Fig.
\ref{FigSTT}, respectively correspond to absence, one and two
divergences in the heat capacity. In the absence of divergence,
the entropy is an increasing function of the temperature.
Interestingly, in case of divergence, there is only one extremum
for the temperature and entropy. Except for this extremum, the
entropy is an increasing function of the temperature. For the case
of two divergences, we can divide the diagram into three regions.
These regions are recognized by two temperatures, $T_{1}$ and
$T_{2}$ which are essentially extrema. Before $T_{1}$ ($T<T_{1}$)
and after $T_{2}$ ($T>T_{2}$), the entropy is an increasing
function of the temperature and for every temperature, there
exists only one entropy. For $T=T_{1},T_{2}$, one can
have two entropies, and interestingly, between these two temperatures ($%
T_{1}<T<T_{2}$), for any temperature, one can find three different
entropies. For all of the plotted diagrams, in the absence of temperature,
one can find a non-vanishing positive entropy. There are several significant
differences between the corrected case and usual one for AdS spacetime: I)
in the case of correction, a minimum was observed for the entropy in certain
non-zero temperature, interestingly, for the non-corrected case, the minimum
for entropy is obtained for the vanishing temperature, II) for the
correction, in case of two divergency in the heat capacity, a cycle was
observed for entropy versus temperature diagram, whereas, this cycle is not
present in the usual non-corrected case, and III) in the correction case,
for every entropy, there exists at least two temperatures except for two
points: one is at the minimum of entropy and the other one, which is more
interesting, is when two branches of $A$ and $B$ meet. This coincidence
between the branches is due to the fact that there is a cycle in correction
case for AdS spacetime. It is worthwhile to mention that for the correction
case, one can point out three distinctive temperatures: $T^{\prime }_{1}$, $%
T^{\prime }_{2}$ and $T_{c}$, where $c$ stand for coincidence. In case of $%
T<T^{\prime }_{1}$ and $T>T^{\prime }_{2}$, for every temperature
there exists one entropy. On the other hand, in case of $T^{\prime
}_{1}<T<T_{c}$ and $T_{c}<T<T^{\prime }_{2} $, each temperature
has three different entropies. Interestingly, in case of
$T=T^{\prime }_{1}, T^{\prime }_{2}, T_{c}$ each temperature is
mapped with two different entropies. The three mentioned points
highlight the significant differences in thermodynamical picture
of the entropy measured by the fluctuation of temperature
comparing to the usual case of without any correction. For further
clarification, we compare the heat capacity of the two cases.

In the previous section, we point out that denominator of the heat
capacity is not affected by the first-order correction. This
indicates that phase transition points (divergences of the heat
capacity) are independent of the first-order correction. On the
contrary, the numerator of the heat capacity was highly correction
dependent. The first result of this dependency is that temperature
and heat capacity do not share the same root anymore (it is
worthwhile to mention that one of the important properties of the
non-correction case is the coincidence between the roots of
temperature and heat capacity). The second important result is
that thermal stability conditions are determined by the value of
correction parameter. In order to highlight this,
we have plotted the diagrams for the AdS case in Figs. \ref{FigCC1} and \ref%
{FigCC2}.

In case of the two divergences in the heat capacity, the effects of
correction could be categorized into three cases with two distinctive
correction parameter, $\alpha_{1}$ and $\alpha_{2}$. I) For $%
\alpha<\alpha_{1}$, the general behavior of the heat capacity is
similar to the one in the non-correction case with slight
modification in place of the root. II) For
$\alpha_{1}<\alpha<\alpha_{2}$, the behavior of heat capacity,
hence stability conditions are modified on a significant level.
Before the smaller divergence, no root is present and the heat
capacity is negative. Between the divergences, a root is observed
for the heat capacity. Before root, the heat capacity is positive
while after it until the larger divergence, it is negative. After
the larger divergence, the heat capacity is positive. Therefore,
the stable regions are between smaller
divergence and root, and after the larger divergence. III) for $%
\alpha_{2}<\alpha$, the root of heat capacity is observed after
larger divergence. Before the smaller divergence and, between
larger divergence and root, the heat capacity is negative which
indicates thermal instability in these regions. Between the
divergences and after the root, the heat capacity is positive and
solutions are thermally stable.

In case of one divergence, the effects of correction are divided
into two groups. I) For small values of correction parameter, the
root of heat capacity is slightly modified to larger values of
horizon radius and the stability conditions are same as those
derived for the non-correction case. II) for sufficiently large
correction parameter, the place of root is changed. It will be
after the divergence. Before divergence and, between divergence
and root, the heat capacity is negative and solutions are
thermally unstable. The only thermal stable state exists after the
root. For the case in which the heat capacity does not have any
divergence, the only effect of the first-order correction is
changing the place of root. In this case, root is an increasing
function of this parameter.

To summarize, we observed that the entropy, as a function of the
temperature, for the correction case has a significant deviation
from the non-correction case. Thermodynamically speaking, the
existence of correction introduces specific properties into
system, for example, these are cyclic/closed diagrams and
existence of a minimum for entropy for non-zero temperature. The
stability conditions were proved to be highly sensitive to the
variation of the correction parameter. While for small values of
it, the stability condition was not modified, for sufficiently
large values, the thermodynamical stability, hence thermodynamical
structure was modified on a significant level. This signals us
with the fact that deviation from the non-correction case with the
corrected one depends on values that correction parameter could
acquire.

\section{Conclusions}

In this paper, we have investigated the effects of the first-order
correction on thermodynamical behavior of the $4$-dimensional
charge black holes in the presence of gravity's rainbow. Through
our study, we have confirmed that modification in entropy (the
first-order logarithmic correction) resulted into modifications in
mass (internal energy), heat capacity and free energy, whereas,
interestingly, no effects were observed for temperature. This
indicated that the effects of consideration of the first-order
correction on entropy and internal energy are on a level which
leaves the temperature unaffected. This situation is very similar
to isothermal process in which internal energy and entropy is
changed while the temperature remains fixed.

One of the significant effects of the correction has been found in
the contexts of heat capacity, more precisely, in the context of
thermal stability of the solutions. Surprisingly, the phase
transition points (divergences of the heat capacity) were not
affected by the presence of the first-order correction while the
number of roots, therefore, the signature of heat capacity was
highly modified to a level of introducing new thermal stability
conditions and presence of multiple stable/unstable phases.
Remarkably, even in case of AdS black holes, the phase transition
between two unstable black holes was observed. This is a
significant difference observed in the case where the first-order
correction is absent.

Generally speaking, the highest contribution of the first-order
correction was on high energy limit of different quantities.
Especially, it was shown that high energy limit of heat capacity
is purely governed by the correction parameter. Moreover, in
evaporation of the black holes, a trace of its existence could be
detected through the remnant of heat capacity in form of
correction parameter. Furthermore, the effects of the first-order
correction could significantly be observed for medium black holes
as well. However, for large black holes, the contributions of this
correction was either absent or not present on a significant
level.

The study of the entropy versus temperature revealed several
interesting facts. First of all, it was shown that dS and AdS
cases have a completely different thermodynamical picture and
limitations. For example, the measurement of entropy as a function
of fluctuation of temperature showed that dS case puts an upper
bound on values that temperature that can be acquired and a cyclic
behavior wasn't observed. On the contrary, for the AdS case, a
lower bound was observed for the entropy, while the cyclic
behavior could be detected for special cases and there were no
upper bound on the temperature. Furthermore, it was observed that
for the cases of two divergences in the heat capacity for AdS
case, the entropy diagrams also recognizes the existence of such
discontinuity in their structure.

The study conducted in this paper confirmed that deviation from
thermodynamical properties of the non-correction case with the
first-order corrected entropy is determined by the correction
parameter, $\alpha $. Especially, it was shown that for the cases
of existence of divergence in the heat capacity, hence phase
transition points, the thermodynamical stability was highly
dependent on the value of correction parameter. This indicates
that thermodynamical structure of the black holes in the presence
of the first-order correction is determined by the value of
correction parameter and the corrected case enjoys larger classes
of possibility for its thermodynamical structure.

In fact, we are faced with a very important question: are all the
classes provided with the first-order correction physically
acceptable? If the answer to this question is positive this means
that correction parameter is not bounded and it can acquire any
value. But then again, we are using the term \textit{correction}
in our analogy. This term implies that we are expecting to see
modifications on a specific level. Therefore, we have to label
some of the possibilities provided by the first-order correction
as \textit{non-physical ones}. This requires finding an
upper/lower limit on values that correction parameter could
acquire. To address this issue, there are following two approaches
that one can employ. I) By considering that the entropy correction
has holographical origins, it is possible to find the bound on
values of the correction parameter by studying superconductivity
properties or diamagnetic/paramagnetic phase transitions and
phases. This enables us to rule out the non-physical cases and
puts bounds on the values of correction parameter. II) The second
method is to use the properties of extended phase space. Recently,
it was shown that by this consideration, one can find
compressibility coefficient and define the speed of sound for
black holes. Using the fact that the speed of sound could not
exceed the speed of light, one is able to recognize the physically
allowed possibilities provided by the first-order correction and
put a limit on correction parameter. The only shortcoming of this
approach is the fact that this method is limited to the cases of
negative cosmological constant. This is due to fact that extended
phase space is constructed by replacing the negative cosmological
constant with thermodynamical pressure. We leave these issues for
future works.

\textbf{Acknowledgements:} S. H. Hendi and B. Eslam Panah thank Shiraz
University Research Council. The work of S. H. Hendi has been supported
financially by the Research Institute for Astronomy and Astrophysics of
Maragha, Iran.

\end{document}